\documentclass[twocolumn,pre,twoside,groupedaddress,floatfix]{revtex4}

\usepackage{graphicx}
\usepackage{color}
\usepackage{placeins}

\usepackage{amsmath}
\usepackage{amssymb}
\usepackage{mathtools}
\usepackage{bbold}

\usepackage{siunitx}
\DeclareSIUnit\Molar{\textsc{M}}

\newcolumntype{K}{>{\centering\arraybackslash$}p{1cm}<{$}}

\renewcommand{\vec}[1]{\boldsymbol{#1}}
\renewcommand{\hat}[1]{\widehat{#1}}
\newcommand{\mat}[1]{\boldsymbol{\underline{\underline{#1}}}}

\def\dps{\displaystyle}

\def\epsilon{\varepsilon}
\def\theta{\vartheta}
\def\rho{\varrho}

\def\diag{\mathop\mathrm{diag}}

\begin{document}

\title{Electrolyte solutions at heterogeneously charged substrates}

\date{March 28, 2018}

\author{Maximilian Mu{\ss}otter}
\email{mussotter@is.mpg.de}
\author{Markus Bier}
\email{bier@is.mpg.de}
\author{S. Dietrich}
\affiliation{
	Max Planck Institute for Intelligent Systems, 
	Heisenbergstr.\ 3,
	70569 Stuttgart,
	Germany
}
\affiliation{
	IV$^{\text{th}}$ Institute for Theoretical Physics, 
	University of Stuttgart,
	Pfaffenwaldring 57,
	70569 Stuttgart,
	Germany
}

\begin{abstract}
	The influence of a chemically or electrically heterogeneous distribution of interaction sites
	at a planar substrate on the number density of an adjacent fluid is
	studied by means of classical density functional theory (DFT). 
	In the case of electrolyte solutions the effect of this heterogeneity is particularly long 
	ranged, because the corresponding relevant length scale is set by the Debye length which is large
	compared to molecular sizes. The DFT used here takes the solvent particles explicitly into account
	and thus captures phenomena, inter alia, due to ion-solvent coupling.
	The present approach provides closed analytic expressions describing the influence of
	chemically and electrically nonuniform walls. 
	The analysis of isolated $\delta$-like interactions, isolated interaction patches, and
	hexagonal periodic distributions of interaction sites reveals a sensitive dependence of
	the fluid density profiles on the type of the interaction, as well as on the size
	and the lateral distribution of the interaction sites.
\end{abstract}

\maketitle

\section{Introduction}

Detailed knowledge of the structure of electrolyte solutions close to solid substrates is of great
importance to numerous research areas and fields of application, ranging from electrochemistry 
\cite{Bagotsky2006, Schmickler2010} and wetting phenomena \cite{Dietrich1988, Schick1990} via 
coating \cite{Wen2017} and surface patterning \cite{Vogel2012, Nee2015} to colloid science 
\cite{Russel1989, Hunter2001} and microfluidics \cite{Lin2011, GalindoRosales2018}. 
The vast majority of models describing fluids in contact with substrates consider the latter as
uniform with respect to the wall-fluid interaction.
This approximation is commonly made partly due to a lack of experimental data on the actual local
properties of the substrate under consideration and partly for the sake of simplicity.
For fluids comprising only \textit{electrically neutral} constituents and \textit{uncharged} walls, assuming uniform
substrates is typically an acceptable approximation because, in the absence of wetting transitions, heterogeneous substrate properties 
influence the fluid only on length scales of the order of the bulk correlation length 
\cite{Andelman1991}, which, not too close to critical points, is of the order of a few molecular 
diameters. 
In contrast, nonuniformities of the \textit{surface charge} density of charged substrates in contact with
dilute electrolyte solutions influence the fluid on the scale of the Debye length, which is much
larger than the size of the molecules. 
Furthermore, the charged sites of substrates, such as mineral surfaces and polyelectrolytes, are
lateral distances apart which are typically comparable with the Debye length of the surrounding
fluid medium \cite{Chen2005, Chen2006, Chen2009}. 
Hence, the assumption, that substrates in contact with electrolyte solutions carry a uniform 
surface charge density, is in general untenable.

In recent years considerable theoretical interest has emerged in the effective interaction
between two heterogeneously charged walls (which typically are the surfaces of colloidal particles)
mediated
by an electrolyte solution \cite{Naji2010, Ben-Yaakov2013, Naji2014, Bakhshandeh2015, Ghodrat2015a,
Ghodrat2015b, Adar2016, Adar2017a, Adar2017b, Ghosal2017, Zhou2017}. 
In contrast to uniform substrates, this effective interaction can lead to lateral forces, in
addition to the common ones in normal direction. 
However, all the studies cited above model the solvent of the electrolyte solution as a 
structureless dielectric continuum.
This approach precludes coupling effects due to a competition between the solvation and the 
electrostatic interaction, which are known to occur in bulk electrolyte solutions \cite{Onuki2004,
Bier2012a, Bier2012b}.
In particular, in the presence of ion-solvent coupling and far away from critical points, correlations of the solvent number densities in a dilute electrolyte solution decay asymptotically on the scale of the
Debye length. Consequently, under such conditions, nonuniformities of the nonelectrostatic solvent-wall interaction
can influence the structure of an electrolyte solution close to a wall and hence the strength and range of the effective
interaction between two parallel plates immersed in an electrolyte medium on a length scale much larger than the molecular size.
This mechanism differs from the one studied in Refs.~\cite{Naji2010, Ghodrat2015a,
Adar2016, Zhou2017}, in which the walls are locally charged but overall charge neutral.

In the present analysis a first step is taken towards a description of the structure of electrolyte solutions
close to chemically and electrically nonuniform walls in terms of all fluid components.
The natural framework for obtaining the fluid structure in terms of number density profiles of solvent
and ion species is classical density functional theory \cite{Evans1979, Evans1990, Evans1992}.
Here, the most simple case of an electrolyte solution, composed of a single solvent species and
a single univalent salt component, is considered far away from bulk or wetting phase transitions.
Moreover, the spatial distribution of nonuniformities of the chemical and electrostatic wall-fluid
interactions can be arbitrary but their strengths are assumed to be sufficiently weak such that a
linear response of the number density deviations from the bulk values is justified.
This setup allows for closed analytic expressions which are used to obtain a first overview of the influence
of ion-solvent coupling on the structure of electrolyte solutions in contact with chemically or
electrically nonuniform walls.
This insight will guide future investigations of more general setups within more sophisticated models.

After introducing the formalism in Sec.~\ref{Sec:Formalism}, selected cases of heterogeneous walls
are discussed in Sec.~\ref{Sec:ResultsDiscussion}.
Due to the linear relationship between the wall nonuniformities and the corresponding number density
deviations from the bulk values, the latter are given by linear combinations of elementary response
features, which are discussed first.
Next, two main cases are studied: wall heterogeneities, which are laterally isotropic
around a certain center and wall heterogeneities, which possess the symmetry of a two-dimensional
lattice; the study of randomly distributed nonuniformities \cite{Naji2010, Naji2014, Bakhshandeh2015, Ghodrat2015a, Ghodrat2015b}
is left to future research.
For both cases various length scale regimes are discussed, which are provided by the bulk correlation
length of the pure solvent, the Debye length, and a characteristic length scale associated with the wall 
nonuniformities.
Conclusions and a summary are given in Sec.~\ref{Sec:ConclusionsSummary}.

\section{Theoretical foundations\label{Sec:Formalism}}

\subsection{Setup}

Here, the influence of a chemically and electrically nonuniform wall on the
fluid density is studied.
In spatial dimension $d=3$ the system consists of an impenetrable
planar wall for $z < 0$ and a fluid for $z > 0$, both parts being macroscopically large.
In the following, the space occupied by the fluid is denoted by $\mathcal{V}:=\{\vec{r}=(x,y,z)
\in\mathcal{A}\times\mathcal{L}\}$; the positions $\vec{r}=(x,y,z)=(\vec{r_\|},z)$ are uniquely 
decomposed into the lateral components $\vec{r_\|}=(x,y)\in\mathcal{A}\subset\mathbb{R}^2$ and the normal
component $z\in\mathcal{L} = [0,L]$ relative to the wall surface at $z=0$. The size $|\mathcal{A}|$ of the wall
and the extent $L$ of the system in normal direction are both assumed to be macroscopically large. The fluid is an electrolyte solution composed of an uncharged solvent (index ``1''), univalent
cations (index ``2''), and univalent anions (index ``3''). 
Two types of interactions between the fluid and the wall are considered: (i) electric
monopoles at the wall surface ($z=0$) and the fluid ions, giving rise to an electrostatic interaction,
(ii) all other contributions, in particular those due to nearest-neighbor-like chemical bonds,
referred to as nonelectrostatic interactions.

\subsection{Density functional theory \label{Sec:DFT}}

We use density functional theory \cite{Evans1979,Evans1990, Evans1992} in order to determine
the equilibrium number density profiles $\vec{\rho}=(\rho_1,\rho_2,\rho_3)$ of the three fluid species. 
Since we focus on length scales larger than the sizes of the fluid particles and on
weak wall-fluid interactions, the following dimensionless density functional within a Cahn-Hilliard-like 
square-gradient approximation \cite{Cahn1958} is considered:
\begin{align}
	\beta\Omega[\vec{\rho}] =&\ \int_\mathcal{V} \mathrm{d}^{3} r~ \bigg[\beta\omega(\vec{\rho}(\vec{r}),\vec{\mu}) + \frac{b}{2}\sum_{j = 1}^3\big(\nabla\rho_j(\vec{r})\big)^2 \nonumber\\&\ + \beta\frac{\epsilon_0\epsilon}{2}\big(\nabla\Psi(\vec{r},[\vec{\rho}])\big)^2\bigg] \nonumber\\&\ - \int_{\mathcal{A}} \mathrm{d}^{2} r_\|~\vec{h}(\vec{r_\parallel})\cdot\vec{\rho}(\vec{r_\parallel},z=0),
	\label{eq:DF}
\end{align}
where $\beta=1/(k_BT)$ is the inverse thermal energy, $\vec{\mu} = (\mu_1,\mu_2,\mu_3)$ are the chemical potentials of the three species, 
$b>0$ is a phenomenological parameter with dimension $[b] = (\text{length})^5$, which can be inferred from
microscopic models (see Sec.~\ref{Sec:ChoiceOfParameters}), $\epsilon_0\approx8.854\times10^{-12}\,\text{As/(Vm)}$
is the vacuum permittivity \cite{Lide1998}, $\epsilon$ is the relative dielectric constant of the fluid, $\Psi(\vec{r},[\vec{\rho}])$ is the 
electrostatic potential at $\vec{r}\in\mathcal{V}$, and $\vec{h}(\vec{r_\|})=
(h_1(\vec{r_\|}),h_2(\vec{r_\|}),h_3(\vec{r_\|}))$ describes the strengths of the nonelectrostatic 
wall-fluid interactions at $\vec{r}=(\vec{r_\|},0)$ for the three species. Note that for the sake of simplicity, the coupling of number density gradients of different particle
types is neglected in Eq. \eqref{eq:DF} (see Sec.~\ref{Sec:ChoiceOfParameters}).
In the present study the bulk state $\vec{\rho}_b=(\rho_{1,b},\rho_{2,b},\rho_{3,b})$ is
considered to be thermodynamically far away from any phase transition so that the local 
contribution $\beta\omega(\vec{\rho})$ of the density functional in Eq.~(\ref{eq:DF}) can be safely
expanded around $\vec{\rho}_b$ up to quadratic order in $\delta\vec{\rho}:=\vec{\rho}-\vec{\rho}_b$ :
\begin{align}
	\beta\omega(\vec{\rho},\vec{\mu}) = \beta\omega(\vec{\rho}_b,\vec{\mu}) + 
	\frac{1}{2}\delta\vec{\rho}\cdot\mat{M}\delta\vec{\rho} 
	\label{eq:local},
\end{align}
where the local part of the interactions between different types of particles is captured
by the real-valued, symmetric, and positively definite $3\times3$-matrix $\mat{M}$ (see, c.f., Eq. \eqref{eq:Mij}).
Furthermore, $\omega(\vec{\rho_b},\vec{\mu}) = -p$ specifies the grand potential density, evaluated for the
equilibrium bulk densities $\vec{\rho_b}$, which equals minus the bulk pressure $p$;
in the following its value is of no importance. For a given equation of state $p(\vec{\rho_b},T)$ the bulk densities $\vec{\rho} = (\rho_{b,1},\rho_{b,2},\rho_{b,3})$ are free parameters
of the model. Finally, the electrostatic potential $\Psi(\vec{r},[\vec{\rho}])$, which enters into Eq.~(\ref{eq:DF}) on a
mean-field level via the electric field energy density, fulfills the Poisson equation
\begin{align}
	-\epsilon_0\epsilon\nabla^2\Psi(\vec{r},[\vec{\rho}]) = e\vec{Z}\cdot\vec{\rho}(\vec{r})
	\label{eq:PE}
\end{align}
for $\vec{r}\in\mathcal{V}$ with the boundary conditions
\begin{align}
	\left.\frac{\partial}{\partial z}\Psi(\vec{r_\|},z,[\vec{\rho}])\right|_{z=0} = 
	-\frac{1}{\epsilon_0\epsilon}\sigma(\vec{r_\|}), 
	\quad
	\Psi(\vec{r_\|},\infty)=0,
	\label{eq:PE_BC}
\end{align}
for $\vec{r_\|}\in\mathcal{A}$, where $\sigma(\vec{r_\|})$ is the surface charge density at the point
$\vec{r}=(\vec{r_\|},0)$ on the wall surface ($z=0$), and $\vec{Z}=(Z_1,Z_2,Z_3)=(0,1,-1)$ denotes the
valences of the fluid species.

The Euler-Lagrange equations, corresponding to the minimum of the density functional specified in Eqs.~(\ref{eq:DF})--(\ref{eq:PE_BC}),
can be written as
\begin{equation}
	b\nabla^2\delta\vec{\rho}(\vec{r}) = \mat{M}\delta\vec{\rho}(\vec{r}) + \beta e \vec{Z}\Psi(\vec{r})\label{eq:ELG1} 
\end{equation}
and
\begin{equation}
	- \frac{1}{4\pi l_B} \nabla^2\beta e \Psi(\vec{r}) = \vec{Z}\cdot\delta\vec{\rho}(\vec{r})\label{eq:ELG2}
\end{equation}
for $\vec{r}\in\mathcal{V}$ with the boundary conditions given by Eq.~(\ref{eq:PE_BC}) and by
\begin{align}
	\frac{\partial}{\partial z}\delta\vec{\rho}(\vec{r_\|},0) = -\frac{1}{b}\vec{h}(\vec{r_\|}),
	\quad
	\delta\vec{\rho}(\vec{r_\|},\infty) = 0
	\label{eq:ELG3}
\end{align}
for $\vec{r_\|}\in\mathcal{A}$, where $l_B=\beta e^2/(4\pi\epsilon_0\epsilon)$ is the Bjerrum
length. 

The linear nature of the Euler-Lagrange equations \eqref{eq:ELG1} and \eqref{eq:ELG2} tells that the quadratic
(Gaussian) approximation of the underlying density functional in Eqs. \eqref{eq:DF} and \eqref{eq:local} corresponds to a linear
response approach. It is widely assumed and in some cases it can be even quantified (see, e.g., the quantitative agreement
between the full and the linearized Poisson-Boltzmann theory in the case that the surface charges are smaller than the saturation value \cite{Russel1989,Bocquet2002})
that for sufficiently weak wall-fluid interactions linear response theory provides quantitatively reliable results.

\subsection{Solution of the Euler-Lagrange equations}

Instead of solving the Euler-Lagrange equations in Eqs.~(\ref{eq:ELG1}) and (\ref{eq:ELG2}) as
differential equations for the profiles $\delta\vec{\rho}$ and $\Psi$ as functions of $\vec{r}=
(\vec{r_\|},z)$, it is convenient first to perform Fourier transformations with respect to the 
lateral coordinates $\vec{r_\|}$. The resulting transformed profiles
\begin{equation}
	\delta\hat{\vec{\rho}}(\vec{q_\|},z) = \int_{\mathcal{A}}\mathrm{d}^2\vec{r_\|}\delta\vec{\rho}(\vec{r_\|},z)\exp(-i\vec{q_\|}\cdot\vec{r_\|})
\end{equation}
and $\hat\Psi$ as functions of 
$\vec{q_\|}=(q_x,q_y)\in\mathbb{R}^2$ and $z\in\mathbb{R}$ can be combined in the four-component 
quantity $\vec{v}(\vec{q_\|},z) = (\delta\hat{\vec{\rho}}(\vec{q_\|},z),\beta e \hat{\Psi}(\vec{q_\|},z))$ 
so that Eqs.~\eqref{eq:ELG1} and \eqref{eq:ELG2} can be written as
\begin{widetext}
\begin{align}
	\underbrace{\left( {\begin{array}{*{4}K}
					b & 0 & 0 & 0 \\
					0 & b & 0 & 0 \\
					0 & 0 & b & 0 \\
					0 & 0 & 0 & -\frac{1}{4\pi l_B} \\
	\end{array} } \right)}_{\dps=:\mat{D}}\vec{v}''= \underbrace{\left( {\begin{array}{cccc}
					\cline{1-3}
					\multicolumn{3}{|c|}{} & Z_1 \\
					\multicolumn{3}{|c|}{\mat{M} + b k^2 \mat{\mathbb{1}} }& Z_2 \\
					\multicolumn{3}{|c|}{} & Z_3 \\\cline{1-3}
					Z_1 & Z_2 & Z_3 & -\frac{k^2}{4\pi l_B} \\
	\end{array} } \right)}_{\dps=:\mat{N}(k)}\vec{v},
	\label{eq:MatELG1}
\end{align}
\end{widetext}
where $k := |\vec{q_\|}| = \sqrt{q_x^2 + q_y^2}$ and $\vec{v}''(\vec{q_\|},z)$ is the second derivative
of $\vec{v}(\vec{q_\|},z)$ with respect to the coordinate $z$ normal to the wall.
Note that the components of $\vec{v}$ are quantities of different dimensions: $[v_1] = [v_2] = [v_3] = 1/\text{length}$ and $[v_4] = (\text{length})^2$.  
This does not allow for the formation of a scalar product of two vectors of the type $\vec{v}=(\delta\widehat{\vec{\rho}},\beta e \widehat{\Psi})$; however,
in the following scalar products will not occur.
Writing $\mat{D} = \mat{T}\,\mat{T}$ with $\mat{T}:=\diag(\sqrt{b},\sqrt{b},\sqrt{b},i\sqrt{1/(4\pi\l_B)})$
(i.e., $\mat{T}$ is a diagonal matrix with these entries), one obtains
\begin{equation}
	\mat{T}\vec{v}''(\vec{q_\|},z) = 
	\underbrace{\mat{T}^{-1}\mat{N}(k)\mat{T}^{-1}}_{\dps=:\mat{H}(k)}\mat{T}\vec{v}(\vec{q_\|},z).
	\label{eq:MatELG2}
\end{equation}
The $4\times4$-matrix $\mat{H}(k)$ is independent of $z$ and it is symmetric but not real-valued, because the bottom
entry of $\mat{T}$ is imaginary. 
$\mat{H}(k)$ is not a normal matrix, i.e., it does not commute with its adjoint matrix $\mat{H}(k)^{\dagger}$, and hence it does not possess an 
\emph{orthogonal} basis composed of eigenvectors.
However, the actual structures of the matrix $\mat{M}$ and of the vector $\vec{Z}$ used below guarantee the existence of
a \emph{nonorthogonal} basis $\{\vec{\Lambda}_1(k),\dots,\vec{\Lambda}_4(k)\}$ of eigenvectors of the matrix 
$\mat{H}(k)$ with respective positive (real-valued) eigenvalues $\lambda_1(k),\dots,\lambda_4(k)\in(0,\infty)$ (see
Appendix~\ref{app:EVEW}).
Expanding the vector $\mat{T}\vec{v}(\vec{q_\|},z)$ in this basis $\{\vec{\Lambda}_1(k),\dots,\vec{\Lambda}_4(k)\}$,
\begin{equation}
	\mat{T}\vec{v}(\vec{q_\|},z) = \sum_{\alpha = 1}^4 A_\alpha(\vec{q_\|},z)~\vec{\Lambda}_\alpha(k),
\end{equation}
leads to Eq.~(\ref{eq:MatELG2}) in the form
\begin{align}
	A_\alpha''(\vec{q_\|},z) = \lambda_\alpha(k)A_\alpha(\vec{q_\|},z)
	\label{eq:DGLA}
\end{align}
with the solution
\begin{align}
	A_\alpha(\vec{q_\|},z) = g_\alpha(\vec{q_\|})~\exp(-\sqrt{\lambda_\alpha(k)}z),
	\label{eq:SolA}
\end{align}
where the second boundary conditions in Eqs.~(\ref{eq:PE_BC}) and (\ref{eq:ELG3}) have been used.
Therefore, the solution of Eq.~\eqref{eq:MatELG2} can be expressed as
\begin{align}
	\vec{v}(\vec{q_\|},z) = 
	\sum_{\alpha = 1}^4 g_\alpha(\vec{q_\|})~\exp(-\sqrt{\lambda_\alpha(k)}z)~\mat{T}^{-1}\vec{\Lambda}_\alpha(k).
	\label{eq:V}
\end{align}
Finally, the first boundary conditions in Eqs.~(\ref{eq:PE_BC}) and (\ref{eq:ELG3}) can be expressed as
\begin{align}
	\vec{v}'(\vec{q_\|},0) = 
	\left(-\frac{1}{b}\hat{\vec{h}}(\vec{q_\|}), -\frac{\beta e}{\epsilon_0\epsilon}\hat\sigma(\vec{q_\|})\right),
	%
	\label{eq:V_BC}
\end{align}
with 
\begin{equation}
	\hat{\vec{h}}(\vec{q_\|}) = \int_{\mathcal{A}}\mathrm{d}^2r_\|\exp(-i\vec{q_\|} \cdot \vec{r_\|})\vec{h}(\vec{r_\|})
\end{equation}
as the Fourier transform of $\vec{h}(\vec{r_\|})$ with respect to the lateral
coordinates $\vec{r_\|}$ and $\hat{\sigma}(\vec{q_\|})$ as the Fourier transform of
$\sigma(\vec{r_\|})$. 
Note that, as for $\vec{v}$, the components of $\vec{v'}$ are quantities of different 
dimensions: $[v'_1] = [v'_2] = [v'_3] = 1/(\text{length})^2$ and $[v_4] = \text{length}$.  
From Eqs.~\eqref{eq:V} and \eqref{eq:V_BC} the coefficients $g_1(\vec{q_\|}),\dots,g_4(\vec{q_\|})$ can be determined.
Note that according to Eqs.~(\ref{eq:V}) and (\ref{eq:V_BC}) the coefficients $g_1(\vec{q_\|}),\dots,g_4(\vec{q_\|})$ and
hence the profiles $\hat{\vec{\rho}}$ and $\hat{\Psi}$ depend linearly on the nonelectrostatic wall-fluid interactions
$\hat{\vec{h}}(\vec{q_\|})$ and the surface charge density $\hat{\sigma}(\vec{q_\|})$.
Such a linear response of the number density profiles inside the fluid to the wall properties requires weak wall-fluid interactions,
which is assumed in the present study and which is consistent with the quadratic form of the density functional in 
Eqs.~(\ref{eq:DF})--(\ref{eq:PE_BC}).

\section{Results and Discussion\label{Sec:ResultsDiscussion}}

\subsection{Choice of parameters\label{Sec:ChoiceOfParameters}}

The present study discusses the influence of the wall-fluid interactions, represented by the 
nonelectrostatic wall-fluid interactions $\hat{\vec{h}}(\vec{q_\|})$ and the surface charge density 
$\hat{\sigma}(\vec{q_\|})$, onto the number density profiles $\vec{\rho}$ in the adjacent fluid.
Applying density functional theory as described in Sec.~\ref{Sec:Formalism} requires knowledge of the 
bulk number densities $\vec{\rho}_b$, the parameter $b$, and the coupling matrix $\mat{M}$ all of which
are bulk quantities or characterize them.

In the bulk local charge neutrality holds, i.e., $\vec{Z}\cdot\vec{\rho}_b=0$.
Hence the equilibrium bulk state is determined by the temperature $T$, the number density 
$\rho_{1,b}$ of the solvent, and the bulk ionic strength $I = \rho_{2,b} = \rho_{3,b}$.

\begin{figure*}
	\includegraphics{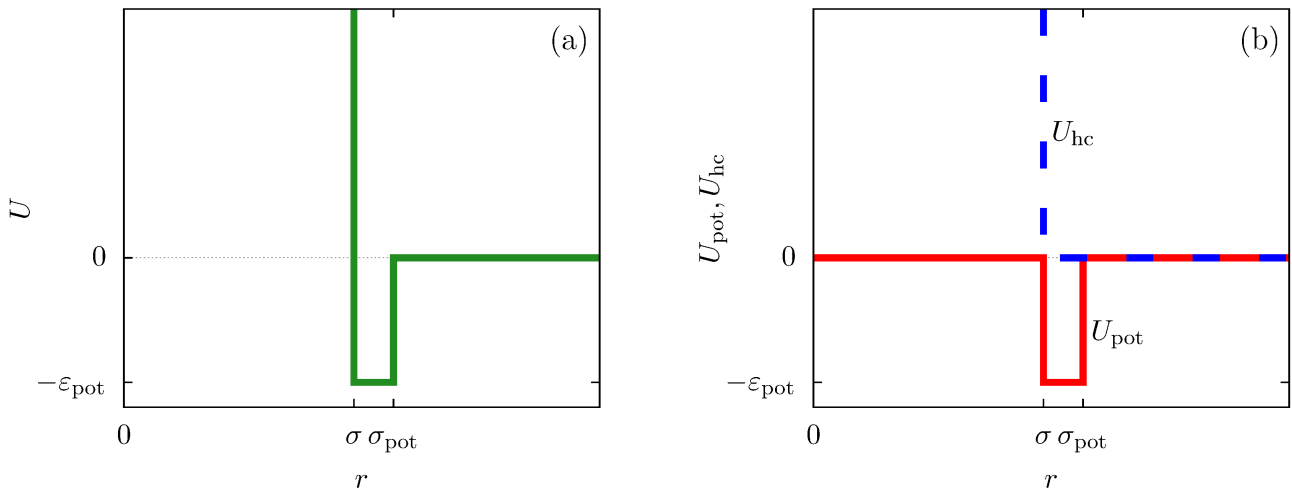}
	\caption{The nonelectrostatic interaction between fluid particles is modeled by a square-well pair
		potential $U(r)$ displayed in panel (a), where $r$ denotes the distance between the
		centers of two spherical particles.
		For $r<\sigma$ a hard core repulsion prevents the overlap of two particles.
		For $r\in(\sigma,\sigma_\text{pot})$ two particles attract each
		other with a constant interaction energy $-\epsilon_\text{pot}<0$.
		At distances $r>\sigma_\text{pot}$ there is no nonelectrostatic interaction.
		Panel (b) sketches the decomposition $U=U_\text{hc}+U_\text{pot}$ of the nonelectrostatic
		interaction potential $U$ according to the scheme due to Barker and Henderson into the
		hard core repulsion $U_\text{hc}$ and the attractive well $U_\text{pot}$, which is used 
		in Sec.~\ref{Sec:ChoiceOfParameters} in order to obtain the parameters entering the Cahn-Hilliard
	square-gradient density functional in Eq.~\eqref{eq:DF}.\label{fig:SquareWell}}
\end{figure*}

In order to obtain expressions for the parameter $b$ and for the coupling matrix $\mat{M}$ in terms of
experimentally accessible quantities, in a first step the pure, ion-free solvent is considered, the 
particles of which interact only nonelectrostatically.
Here this nonelectrostatic interaction between solvent particles at a distance $r$ is modeled by a square-well pair
potential $U(r)$ as displayed in Fig.~\ref{fig:SquareWell}(a).
At small distances $r<\sigma$ a hard core repulsion prevents the overlap of two particles. At
intermediate distances $r\in(\sigma,\sigma_\text{pot})$ two particles attract each other with an interaction energy
$-\epsilon_\text{pot}$, and at distances $r>\sigma_\text{pot}$ the nonelectrostatic interaction vanishes.
According to the scheme due to Barker and Henderson \cite{Hansen1986}, the interaction potential $U$ can be decomposed as 
$U=U_\text{hc}+U_\text{pot}$ into the hard core repulsion $U_\text{hc}$ and the attractive well 
$U_\text{pot}$ (see Fig.~\ref{fig:SquareWell}(b)).
The microscopic density functional $\Omega^\text{mic}_1[\rho_1]$ for the pure solvent (species 
$1$) in the bulk can be approximated by the expression
\begin{align}
	\beta\Omega^\text{mic}_1[\rho_1] = \beta\Omega^\text{hc}_1[\rho_1] + \beta F^\text{ex,pot}[\rho_1].
	\label{eq:Omegamic}
\end{align}
The contribution $\Omega_1^{\text{mic}}$ (here within local density approximation (LDA)) is due to the reference
system governed solely by the hard core interaction $U_\text{hc}$:
\begin{align}
	\beta\Omega^\text{hc}_1[\rho_1] =
	\int_\mathcal{V} \mathrm{d}^{3} r~
	&\left[\rho_1(\vec{r})\left(\ln(\rho_1(\vec{r})\Lambda_1^3)-1-\beta\mu_1\right)\right.\nonumber\\
	&\left.+\beta f^\text{ex,hc}(\rho_1(\vec{r}))\right].
	\label{eq:LDAmic}
\end{align}
Here, contributions of external potentials are neglected, because they do not contribute to the bulk parameters $b$ and $\mat{M}$.
The second term on the rhs of Eq. \eqref{eq:Omegamic} is (within random phase approximation (RPA) \cite{Evans1979}) the
excess free energy functional due to the square-well attractive interaction $U_\text{pot}$: 
\begin{align}
	\beta F^\text{ex,pot}[\rho_1] =
	\frac{1}{2}\int_\mathcal{V} \mathrm{d}^{3} r\int_\mathcal{V} \mathrm{d}^{3} r'~
	\beta U_\text{pot}(\vec{r}-\vec{r'})\rho_1(\vec{r})\rho_1(\vec{r'}).
	\label{eq:RPAmic}
\end{align}
In Eq.~\eqref{eq:LDAmic} $\Lambda_1$ is the thermal de Broglie wave length, $\mu_1$ denotes the chemical
potential of species 1, and $f^\text{ex,hc}(\rho_1)$ is the excess free energy per volume of the reference
system governed by the hard core interaction $U_\text{hc}$.

Following Cahn and Hilliard \cite{Cahn1958}, Eq. \eqref{eq:RPAmic} can be approximated by a gradient expansion:
\begin{align}
	\beta F^\text{ex,pot}[\rho_1] \simeq
	\int_\mathcal{V} \mathrm{d}^{3} r~\left[\frac{K_0}{2}\left(\rho_1(\vec{r})\right)^2 - 
	\frac{K_2}{12}\left(\nabla\rho_1(\vec{r})\right)^2\right]
	\label{eq:CahnHilliard}
\end{align}
with the $m$-th moment of the pair potential $U_\text{pot}$ in units of $k_BT$, 
\begin{align}
	K_m = \int_{\mathbb{R}^3}\mathrm{d}^3 r~|\vec{r}|^m\beta U_\text{pot}(|\vec{r}|),
	\label{eq:Km}
\end{align}
which, for the present form of $U_\text{pot}$, leads to
\begin{align}
	K_0 &= -\frac{4\pi}{3}\beta\epsilon_\text{pot}\left(\sigma_\text{pot}^3 - \sigma^3\right)<0,
	\notag\\
	K_2 &= -\frac{4\pi}{5}\beta\epsilon_\text{pot}\left(\sigma_\text{pot}^5 - \sigma^5\right)<0.
	\label{eq:K02}
\end{align}
From Eq.~\eqref{eq:CahnHilliard} one obtains the gradient expansion of $\beta\Omega^\text{mic}_1[\rho_1]$
in Eq.~\eqref{eq:Omegamic}:
\begin{align}
	\beta\Omega^\text{mic}_1[\rho_1] \simeq
	\int_\mathcal{V} \mathrm{d}^{3} r~\left[\beta\omega^\text{loc}_1(\rho_1(\vec{r}), \mu_1)
	- \frac{K_2}{12}\left(\nabla\rho_1(\vec{r})\right)^2\right]
	\label{eq:OmegamicGE}
\end{align}
with the local contribution
\begin{align}
	\beta\omega^\text{loc}_1(\rho_1,\mu_1) =
	&\rho_1\left(\ln(\rho_1\Lambda_1^3)-1-\beta\mu_1\right)\nonumber\\
	+&\beta f^\text{ex,hc}(\rho_1) + \frac{K_0}{2}\rho_1^2.
	\label{eq:omegaloc}
\end{align}
The comparison of Eq.~\eqref{eq:OmegamicGE} with Eq.~\eqref{eq:DF} renders an expression for
the parameter $b$ in terms of parameters of the interaction potential $U$ (see Fig. \ref{fig:SquareWell}):
\begin{align}
	b 
	= -\frac{K_2}{6}
	= \frac{2\pi}{15}\beta\epsilon_{\text{pot}}\left(\sigma_{\text{pot}}^5 - \sigma^5\right).
	\label{eq:b}
\end{align}
By expanding $\beta\omega^\text{loc}_1(\rho_1,\mu_1)$ up to quadratic order in the density
deviation $\delta\rho_1 = \rho_1-\rho_{1,b}$ from the equilibrium bulk density $\rho_{1,b}$, which
is a solution of the Euler-Lagrange equation
\begin{align}
	0 = \frac{\partial\ (\beta\omega^\text{loc}_1)}{\partial\rho_1}(\rho_{1,b},\mu_1),
	\label{eq:ELGbulk}
\end{align}
one obtains
\begin{widetext}
\begin{align}
	\beta\omega^\text{loc}_1(\rho_1,\mu_1) 
	&\simeq
	\beta\omega^\text{loc}_1(\rho_{1,b},\mu_1) +\frac{1}{2}\frac{\partial^2\ (\beta\omega^\text{loc}_1)}{{(\partial\rho_1)}^2}(\rho_{1,b},\mu_1)
	(\delta\rho_1)^2 \notag\\
	&= 
	\beta\omega^\text{loc}_1(\rho_{1,b},\mu_1)+\frac{1}{2}\left(\frac{1}{\rho_{1,b}} + 
	\frac{\mathrm{d}^2\ (\beta f^\text{ex,hc}(\rho_{1,b}))}{{(\mathrm{d}\rho_{1,b})}^2} + K_0\right)
	(\delta\rho_1)^2.
	\label{eq:omegalocexpansion}
\end{align}
\end{widetext}
The comparison with Eq.~\eqref{eq:local} leads to the matrix element
\begin{align}
	M_{11} 
	= \frac{1}{\rho_{1,b}} + 
	\frac{\mathrm{d}^2\ (\beta f^\text{ex,hc}(\rho_{1,b}))}{{(\mathrm{d}\rho_{1,b})}^2} + 
	K_0
	\label{eq:M11}
\end{align}
of the matrix $\mat{M}$, where the first term on the rhs stems from the ideal gas
contribution of the solvent particles. The argument $\rho_{1,b}$ of the second term, which is due to the
hard core interaction $U_\text{hc}$, is a measure of the packing fraction $\eta=\pi\rho_{1,b}\sigma^3/6$.

The analogue of Eq. \eqref{eq:OmegamicGE} for the nonelectrostatic contribution of all \emph{three} particle species is given
by the first line of Eq. \eqref{eq:DF} with the local contribution (compare Eq. \eqref{eq:omegaloc}) 
\begin{align}
	\beta\omega^\text{loc}(\vec{\rho},\vec{\mu}) &=
	\sum_{i=1}^{3}\rho_i\left(\ln(\rho_i\Lambda_i^3)-1-\beta\mu_i\right)\nonumber\\
	&+\beta f^\text{ex,hc}(\rho^{\text{tot}}) + \sum_{i,j=1}^{3}\frac{K_0}{2}\rho_i\rho_j,
	\label{eq:omegalocallparticles}
\end{align}
where $\rho^{\text{tot}} = \rho_{1} + \rho_{2} + \rho_{3}$ denotes the total number density. Note that Eq. \eqref{eq:omegalocallparticles}
	assumes, that all interactions among the species are the same (see Eq. \eqref{eq:Km}). This implies that the last term in Eq.
\eqref{eq:omegalocallparticles} takes the form $\frac{K_0}{2}(\rho^{\text{tot}})^2$.
By expanding $\beta\omega^\text{loc}(\vec{\rho},\vec{\mu})$ up to quadratic order in the density
deviations $\delta\vec{\rho} = \vec{\rho}-\vec{\rho}_b$ from the equilibrium bulk densities $\vec{\rho}_{b}$ one
finally finds (see the steps leading to Eq. \eqref{eq:M11})
\begin{align}
	M_{ij} 
	&= \frac{\delta_{ij}}{\rho_{i,b}} + 
	\frac{\mathrm{d}^2\ (\beta f^\text{ex,hc}(\rho^{\text{tot}}_b))}{{(\mathrm{d}\rho^{\text{tot}}_b)}^2} +
	K_0
	\notag\\
	&= \frac{\delta_{ij}}{\rho_{i,b}} +
	\frac{\mathrm{d}^2\ (\beta f^\text{ex,hc}(\rho^{\text{tot}}_b))}{{(\mathrm{d}\rho^{\text{tot}}_b)}^2} -
	\frac{4\pi}{3}\beta\epsilon_\text{pot}\left(\sigma_\text{pot}^3 - \sigma^3\right),\notag\\
	&\quad i\in\{1,2,3\},
	\label{eq:Mij}    
\end{align}
where $\rho^{\text{tot}}_b = \rho_{1,b} + \rho_{2,b} + \rho_{3,b} = \rho_{1,b} + 2I$ denotes the total number density in
the bulk.
In the present study the hard core excess free energy per volume $f^\text{ex,hc}(\rho_b)$ is chosen
as the one corresponding to the Carnahan-Starling equation of state \cite{Hansen1986}:
\begin{align}
	\beta f^\text{ex,hc}(\rho^{\text{tot}}_b) = \rho^{\text{tot}}_b\frac{\eta(4-3\eta)}{(1-\eta)^2}
	\label{eq:Carnahan-Starling}
\end{align}
here with the packing fraction $\eta = \pi\rho^{\text{tot}}_b\sigma^3/6$.

Accordingly, from Eq.~\eqref{eq:omegaloc} one obtains the following equation of state of the \emph{pure} solvent:
\begin{align}
	\beta p(\rho_{1,b}) = \rho_{1,b}\frac{1+\eta+\eta^2-\eta^3}{(1-\eta)^3} + \frac{K_0}{2}(\rho_{1,b})^2.
	\label{eq:EOSsolvent}
\end{align}
Its derivative with respect to the number density $\rho_{1,b}$, using the relation
$\partial p/\partial\rho_{1,b}=1/(\kappa_T\rho_{1,b})$ with the isothermal compressibility $\kappa_T$,
yields	
\begin{align}
	\frac{\beta}{\kappa_T\rho_{1,b}} =
	\frac{1+4\eta+4\eta^2-4\eta^3-\eta^4}{(1-\eta)^4} + K_0\rho_{1,b}.
	\label{eq:compress}
\end{align}
As an example we consider water at room temperature $T=\SI{300}{\kelvin}$ and ambient pressure 
$p=\SI{1}{\bar}$ (which corresponds to the number density $\rho_{b,1} = \SI{55.5}{\Molar} \approx 
\SI{33.3}{\per\nano\meter\cubed}$ and the isothermal compressibility 
$\kappa_T=\SI{4.5e-10}{\per\pascal}$ \cite{Lide1998}) with relative dielectric constant $\epsilon = 80$,
i.e., with Bjerrum length $l_B = \SI{0.7}{\nano\meter}$, and with a univalent salt of ionic strength 
$I=\SI{1}{\milli\Molar} \approx \SI{6e-4}{\per\nano\meter\cubed}$.
The strength of hydrogen bonds, which generate the dominant attractive interaction 
contribution, is of the order of $\epsilon_{\text{pot}} \approx \SI{20}{\kilo\joule\per\mol}\approx 
8\,k_BT$ \cite{Naik2000,Lide1998}.
Using these data, one obtains from Eqs.~\eqref{eq:EOSsolvent} and \eqref{eq:compress} the bulk packing
fraction $\eta\approx0.44$ as well as $\sigma=\SI{2.9}{\angstrom}$ and 
$\sigma_\text{pot}=\SI{3.4}{\angstrom}$.
In the following the Debye length
\begin{equation}
	\frac{1}{\kappa} = \sqrt{\frac{1}{8\pi l_B I}}
	\label{eq:DEBYE}
\end{equation}
is used as length scale, which, for the present choice of parameters, is $1/\kappa \approx 
\SI{10}{\nano\meter}$.

In the case of a pure solvent ($\delta\rho_2 = \delta\rho_3 = \Psi = 0$),
in the bulk the density two-point correlation function $G(\vec{r}_1,\vec{r}_2) = \bar{G}(\vec{r}_1 - \vec{r}_2)$
fulfills an equation similar to Eq. \eqref{eq:ELG1}:
\begin{equation}
	b\nabla^2\bar{G}(\vec{r}) = M_{11}\bar{G}(\vec{r}).
	\label{eq:G}
\end{equation}
Note that the similarity between Eqs. \eqref{eq:ELG1} and \eqref{eq:G} is due to the asymptotic proportionality between density deviations and
two-point correlation functions (Yvon equation) \cite{Hansen1986}.
From Eq. \eqref{eq:G}, one can readily infer the relation 
\begin{equation}
	\xi = \sqrt{\frac{b}{M_{11}}}
	\label{eq:XI}
\end{equation}
for the solvent bulk correlation length, which characterizes the exponential decay of $\bar{G}(\vec{r})$. For the present
choice of parameters, one has $\xi \approx \SI{1.3}{\angstrom}$ so that $\kappa\xi \approx 0.013$.

\subsection{X-ray scattering\label{Sec:Reflectivity}}

In the following subsections the Fourier transforms $\delta\widehat{\vec{\rho}}=(\delta\widehat{\rho}_1,\delta\widehat{\rho}_2,
\delta\widehat{\rho}_3)$ of the profiles of the density deviations as functions of the lateral wave vector $\vec{q_\|}$ and
of the normal distance $z$ from the wall are discussed in detail.
However, from the experimental point of view, it is challenging to directly obtain the $z$-dependence of the density
profiles.
One of such direct methods is total internal reflection microscopy (TIRM) \cite{Walz1997} in the context of the structure of colloidal
suspensions close to (optically transparent) substrates.
In contrast, for molecular fluids, as the ones considered here, such direct methods are not
available and one has to resort to, e.g., X-ray scattering techniques \cite{AlsNielsen2001,Dietrich1995}.
As X-rays are predominantly scattered by the electrons of the fluid molecules one has to consider the electron number density
\begin{align}
	\rho^e(\vec{r_\|},z) = 
	\sum_{j=1}^3 N_j\rho_j(\vec{r_\|},z) =
	\left\{\begin{array}{ll}
			\rho^e_b + \delta\rho^e(\vec{r_\|},z) & \text{, $z>0$} \\
			0                                     & \text{, $z<0$}
	\end{array}\right.
	\label{eq:rhoe}
\end{align}
for $\vec{r_\|} \in \mathcal{A}$ with the number $N_j$ of electrons per molecule of particle species $j\in\{1,2,3\}$,
the bulk electron density $\rho^e_b=\sum_{j=1}^3N_j\rho_{j,b}$, and the deviation
$\delta\rho^e = \rho^e-\rho^e_b = \sum_{j=1}^3N_j\delta\rho_j$ of the electron number density from its bulk value.
The X-ray scattering signal for scattering vector $\vec{q}=(\vec{q_\|},q_z)$ is proportional to 
$\left|\widehat{\widehat{\rho^e}}(\vec{q_\|},q_z)\right|^2$ with the double Fourier transform
$\widehat{\widehat{\rho^e}}$ of the electron number density profile $\rho^e$ in both the lateral and the normal direction \cite{Hansen1986}.

\begin{figure*}
	\includegraphics{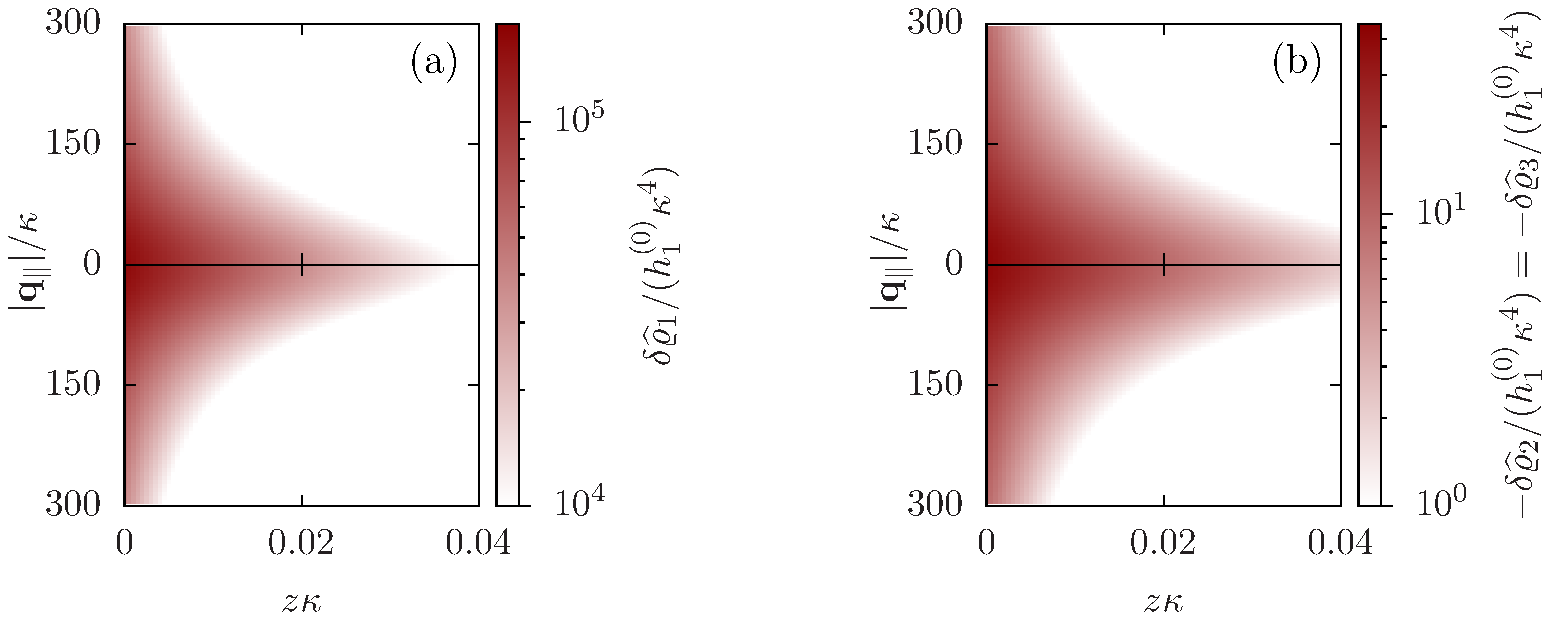}
	\caption{Density distribution $\delta\hat{\rho}_1(\vec{q}_\|,z)$ of the solvent (panel (a)) and of the ions $\delta\hat{\rho}_2(\vec{q}_\|,z)=\delta\hat{\rho}_3(\vec{q}_\|,z)$ (panel (b)) as function of the distance $z$ from the wall and of the absolute value of the lateral Fourier wave vector $\vec{q}_\|$ in units of the inverse Debye length $\kappa$ (see Eq. \eqref{eq:DEBYE}). The plane $z=0$ is given by the positions of the fluid particle centers when the surface-to-surface distance between the hard wall and the hard particles vanishes. The data correspond to the boundary condition $\vec{v}^\prime = -\frac{h_1^{(0)}}{b} (1,0,0,0)$ (see Eqs. \eqref{eq:V_BC} and \eqref{eq:BC1000Eq2}). The physical situation corresponding to this boundary condition is an attraction $h_1(\vec{r_\|}) = h_1^{(0)} \delta(\vec{r_\|})$ (see Eq. \eqref{eq:BC1000EqReal}) of the solvent particles by the wall located at the origin of the wall. Concerning the remaining relevant parameters see Sec. \ref{Sec:ChoiceOfParameters}.\label{fig:density1000}}
\end{figure*}

For common specular X-ray reflectivity measurements, i.e., for $\vec{q_\|}=0$, the normalized intensity reflected as function of the
normal wave number $q_z$ is given by \cite{AlsNielsen2001,Dietrich1995}
\begin{align}
	\frac{R(q_z)}{R^\text{F}(q_z)} = 
	\left|1 + \frac{iq_z\delta\widehat{\widehat{\rho^e}}(\vec{q_\|} = 0,q_z)}{|\mathcal{A}|\rho^e_b}\right|^2, 
	\label{eq:R_RF}
\end{align}
where $R^\text{F}(q_z)$ denotes the Fresnel reflectivity of an ideal, step-like planar interface \cite{Jackson1999}, and where 
the notation $\delta\widehat{\widehat{\rho^e}} := \sum_{j=1}^3N_j\delta\widehat{\widehat{\rho_j}}$ with
\begin{align}
	\delta\widehat{\widehat{\rho_j}}(\vec{q_\|},q_z) = \int\limits_0^\infty\mathrm{d}z~\delta\widehat{\rho}(\vec{q_\|},z)\exp(-iq_zz)
	\label{eq:delta2hatrho}
\end{align}
has been used.
Moreover, off-specular diffuse X-ray scattering ($\vec{q_\|}\not=0$) at grazing incidence
(GIXD, $\mathrm{Im}\ q_z\not=0$) yields scattering intensities which are proportional to
$\left|\delta\widehat{\widehat{\rho^e}}(\vec{q_\|},q_z)\right|^2$ \cite{Dietrich1995}.
Hence, as the double Fourier transforms $\delta\widehat{\widehat{\rho_j}}$ in Eq.~(\ref{eq:delta2hatrho}) of the density deviation profiles 
$\delta\rho_j$ are of direct experimental relevance, they will be discussed in the following in parallel to the single Fourier transforms
$\delta\widehat{\rho}_j$. Note that due to $\delta\widehat{\widehat{\rho_i}}\in\mathbb{C}$, in Figs. \ref{fig:DensitiesElementaryVectors1000}
and \ref{fig:DensitiesElementaryVectors0001} its absolute value is shown. 

\subsection{Basis vectors of boundary conditions\label{Sec:ResultsBasisVector}}

As mentioned above, the linear nature of the relationship between wall nonuniformities and the resulting number density deviations leads to 
the possibility of describing the latter in terms of linear combinations of elementary response patterns. These elementary response patterns
correspond to four basis vectors, e.g., $(1,0,0,0),(0,1,0,0), (0,0,1,0), \text{ and }(0,0,0,1)$, which span the four-dimensional space of boundary
conditions $\vec{v}'(\vec{q_\|},0)$ in Eq. \eqref{eq:V_BC}. Therefore, as a first step to study the influence of wall inhomogeneities onto the fluid,
these four distinct boundary condition vectors $\vec{v}'(\vec{q_\|},0)$ are studied. The first one of these vectors is given by
\begin{equation}
	\vec{v}'(\vec{q_\|},0) = -\frac{h_1^{(0)}}{b}(1,0,0,0),
	\label{eq:BC1000Eq2}
\end{equation}
which requires (see Eq. \eqref{eq:V_BC})
\begin{align}
	\hat{h_1}(\vec{q_\|}) &= h_1^{(0)},\nonumber \\
	\hat{h_2}(\vec{q_\|}) = \hat{h_3}(\vec{q_\|}) & = \hat{\sigma}(\vec{q_\|}) = 0,
	\label{eq:BC1000Eq}
\end{align}
and which in real space corresponds to the boundary condition
\begin{align}
	h_1(\vec{r_\|}) & = h_1^{(0)}~\delta(\vec{r}_\|),\nonumber \\
	h_2(\vec{r_\|}) = h_3(\vec{r_\|}) & = \sigma(\vec{r_\|}) = 0.
	\label{eq:BC1000EqReal}
\end{align}
This boundary condition corresponds to an attractive, $\delta$-like interaction of the wall with the solvent located at the origin.
Solving the Euler-Lagrange equations for this boundary condition, one finds the density distribution $\delta{\hat{\rho_1}}(\vec{q_\|},z)$ for the solvent
and $\delta\hat{\rho}_2(\vec{q}_\|,z)=\delta\hat{\rho}_3(\vec{q}_\|,z)$ for the ions, as shown in Figs. \ref{fig:density1000} (a) and \ref{fig:density1000} (b), respectively. 
Since the boundary condition corresponds to a constant in Fourier space, the density distribution $\vec{v}(\vec{q_\|},z)$ depends on $k = |\vec{q_\|}|$ only. Actually, the solution $\vec{v}(\vec{q_\|},z)$
for this system is proportional to the first column of the Green's function, which is a $4\times4$-matrix, of the differential operator corresponding to Eqs. \eqref{eq:ELG1} and \eqref{eq:ELG2}.

\begin{figure*}
	\includegraphics[scale=0.9]{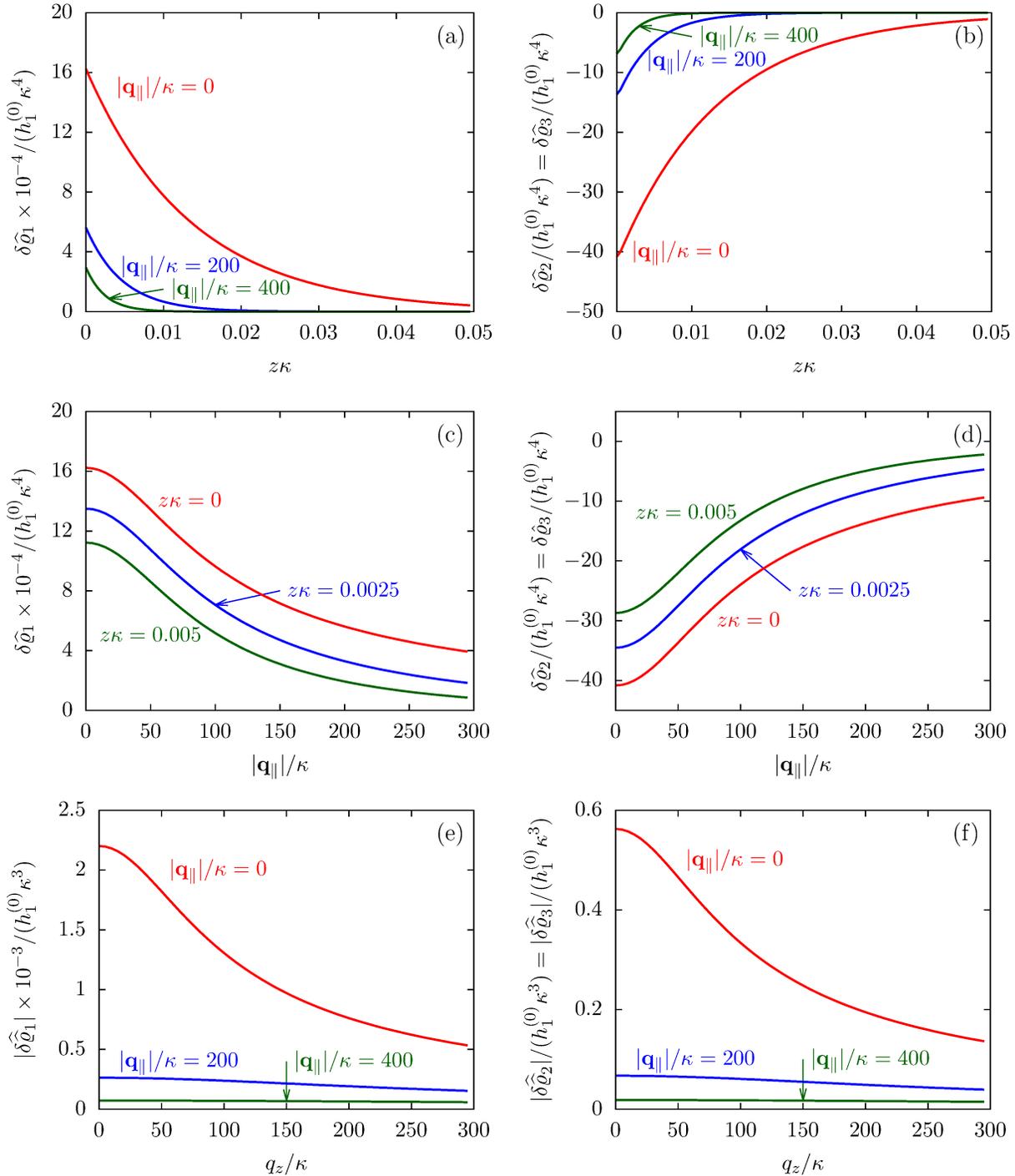}
	\caption{Density profiles of the solvent (left column, panels (a), (c), and (e)) and of the ions (right column, panels (b), (d), and (f)) as functions of the normal distance $z$ from the wall (top row, panels (a) and (b)), of the absolute value of the lateral wave vector $|\vec{q_\|}|$ (middle row, panels (c) and (d)), and of the wave number $q_z$ in normal direction (bottom row, panels (e) and (f), with $h_1^{(0)}\kappa^3$ being dimensionless) in corresponding units of the Debye length $1/\kappa$ and the inverse Debye length, respectively (see Eq. \eqref{eq:DEBYE}). Note that due to $\delta\widehat{\widehat{\rho_i}}\in \mathbb{C}$, in panels (e) and (f) the absolute values are shown. In each graph, there are three profiles shown corresponding to three values of the other relevant variable. Therefore the profiles correspond to cuts through Figs. \ref{fig:density1000} (a) and (b) at various positions and in different directions. In this case the boundary condition is $\vec{v}^\prime = -\frac{h_1^{(0)}}{b}(1,0,0,0)$ (see Eqs. \eqref{eq:V_BC} and \eqref{eq:BC1000Eq2}), corresponding to a $\delta$-like nonelectrostatic attraction of the solvent particles at the origin of the wall (see Fig. \ref{fig:density1000} and Eq. \eqref{eq:BC1000EqReal}). The graphs show, that the density deviations of the ions are proportional to the ones of the solvent, although different in sign. Since only the solvent particles are attracted by the wall, it is favorable for the system to increase their density close to the wall. However, due to the hard core nature of the particles and the equality of the interparticle attraction for all pairs of particles, the increase of solvent particles leads to an extrusion of ionic particles, leading to decreased ion densities at the wall. However, the density deviations of the ions are much weaker. For the remaining relevant parameters see Sec. \ref{Sec:ChoiceOfParameters}.\label{fig:DensitiesElementaryVectors1000}}
\end{figure*}

Figure \ref{fig:density1000} illustrates that for fixed $|\vec{q_\|}|$ the density deviations from the bulk value increase for smaller normal distances from the wall and, for fixed $z$, also upon decreasing 
the absolute value of the lateral wave vector $|\vec{q_\|}|$. The behavior with respect to the normal distance from the wall can be anticipated, because the effect of the interaction between the wall
and the fluid is expected to decrease with increasing distance from the wall. Moreover, also the behavior with respect to $|\vec{q_\|}|$ is
as expected, because a strong attraction at the origin leads to a radially decreasing density deviation, which in Fourier space corresponds to a maximum at the origin.
In order to allow for a quantitative analysis of the behavior of the density deviations, Fig. \ref{fig:DensitiesElementaryVectors1000} shows various cuts through the data of
Fig. \ref{fig:density1000} along several lines.

Figures \ref{fig:DensitiesElementaryVectors1000} (a), (c), and (e) show the density profiles for the solvent and Figs. \ref{fig:DensitiesElementaryVectors1000} (b), (d), and (f)
the ones of the positive ions, which in this case are the same as the profiles for the negative ions.
This equivalence is due to the nature of the boundary conditions in this special case, which in real space lead primarily to an increased solvent density close
to the origin at the wall. The ions, however, react only indirectly via the solvent, with which both ion types interact in the same way. Since the solvent particles get
attracted by the wall, it is favorable to increase their density close to the wall. Due to the hard core nature of the particles, the space occupied
by the solvent particles is blocked for the ions. Since the solvent is attracted by the wall and the interparticle attraction is the same for all pairs of particles, 
this leads to an extrusion of the ions in favor of an increased number of solvent particles.
Figures \ref{fig:DensitiesElementaryVectors1000} (a) and (b) show the density deviations as function of the normal distance $z$ from the wall for three values of $|\vec{q}_\||$, i.e.,
Figs. \ref{fig:DensitiesElementaryVectors1000} (a) and (b) correspond to horizontal cuts through Figs. \ref{fig:density1000} (a) and (b), respectively. For fixed $|\vec{q_\|}|$,
as in Figs. \ref{fig:density1000} (a) and (b), Figs. \ref{fig:DensitiesElementaryVectors1000} (a) and (b) clearly show an exponential decay of the density deviation for increasing distances from the wall.
In contrast, Figs. \ref{fig:DensitiesElementaryVectors1000} (c) and (d) show vertical cuts through Figs. \ref{fig:density1000} (a) and (b), i.e., density profiles as functions of the absolute
value of the lateral wave number $|\vec{q_\|}|$ for three normal distances $z$ from the wall. 
The dependence of these profiles on the absolute value $|\vec{q_\|}|$ of the lateral wave
vector $\vec{q_\|}$ implies a laterally isotropic decay of the density deviations in real space. 
The third pair of graphs, Figs. \ref{fig:DensitiesElementaryVectors1000} (e) and (f), shows the Fourier transforms of the density profiles of Figs. \ref{fig:DensitiesElementaryVectors1000} (a) and (b), 
being additionally Fourier-transformed with respect to the normal direction $z$, which leads to the Fourier transforms $\delta\widehat{\widehat{\rho}}(\vec{q_\|},q_z)$ in terms of the lateral wave vector
$\vec{q_\|}$ and the normal wave number $q_z$, respectively. 
All curves in Figs.~\ref{fig:DensitiesElementaryVectors1000} (c)--(f) exhibit a Lorentzian shape
as functions of $|\vec{q_\|}|$ and $q_z$, respectively. 
These Lorentzian curves in Fourier space correspond to exponential decays in real space in lateral or normal direction.
The curves in Figs. \ref{fig:DensitiesElementaryVectors1000}(c) and (d) show widths of half height which decrease with increasing normal distance $z$,
i.e., the lateral decay length in real space increases with increasing distance from the wall. This implies that the density distribution broadens upon moving away
from the source of the perturbation. The curves in Figs. \ref{fig:DensitiesElementaryVectors1000}(e) and (f)
exhibit widths of half height which increase with the lateral wave number $|\vec{q_\|}|$, i.e., the normal decay length in real space decreases with increasing lateral
wave number. Consequently, the range of influence of rapidly varying modes of wall heterogeneities onto the fluid is shorter than that of slowly varying modes. This relationship 
can also be inferred from Figs. \ref{fig:DensitiesElementaryVectors1000}(a) and (b)).
From the above discussions and from Fig. \ref{fig:DensitiesElementaryVectors1000} one can conclude, 
that the response of all species to a simple attraction of nonelectrostatic type is the same up to a proportionality factor. This is confirmed by studying in addition the boundary conditions
$\vec{v}^\prime = -\frac{h_{2}^{(0)}}{b}(0,1,0,0)$ and $\vec{v}^\prime = -\frac{h_{3}^{(0)}}{b}(0,0,1,0)$; these results are not shown here.

\begin{figure}
	\includegraphics[scale=0.8]{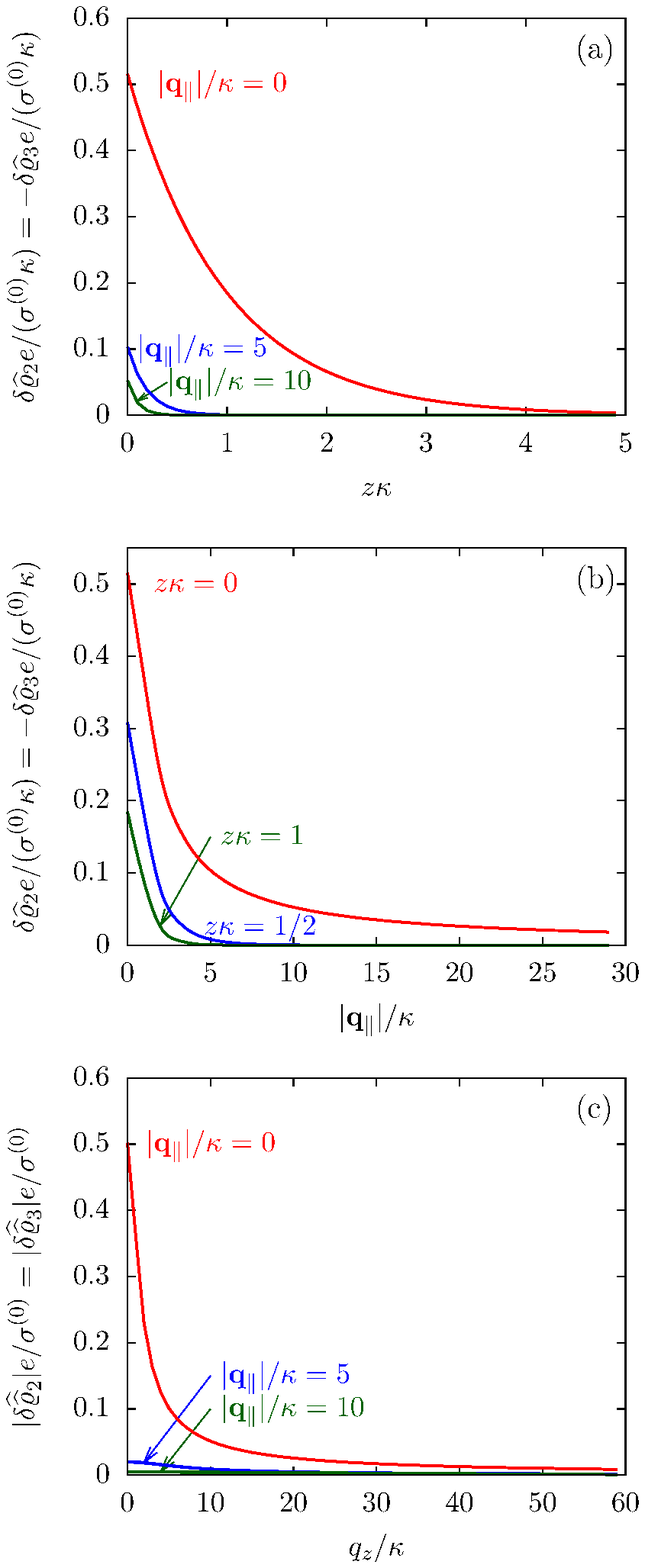}
	\caption{Density profiles of the ions as functions of the normal distance from the wall (a), of the absolute value of the lateral wave vector $|\vec{q_\|}|$ (b), and of the wave number in normal direction (c). Note that due to $\delta\widehat{\widehat{\rho_i}}\in\mathbb{C}$, in panel (c) the absolute value is shown. Each panel shows the profiles for three values of the other relevant variable. These profiles are cuts of the corresponding data (analogous to Fig. \ref{fig:density1000}) along various directions. Here, the boundary condition is given by $\vec{v}^\prime = -\frac{\beta e \sigma^{(0)}}{\epsilon_0\epsilon} (0,0,0,1)$ (see Eqs. \eqref{eq:V_BC} and \eqref{eq:BC0001Eq2}), which corresponds to a $\delta$-like surface charge at the origin in real space (see Eq. \eqref{eq:BC0001Eq}). The profiles for the solvent are not shown, because the deviations linked to the two types of ions cancel out, $\delta\widehat{\rho}_1 + \delta\widehat{\rho}_2 = 0$, leaving the density of the solvent unchanged as if there were no ions. In comparison with Fig. \ref{fig:DensitiesElementaryVectors1000}, the profiles in (a) decay much slower on the scale of the Debye length $1/\kappa$ (see Eq. \eqref{eq:DEBYE}) instead of on the scale of the much shorter bulk correlation length $\xi$ (see Fig. \ref{fig:DensitiesElementaryVectors1000}(b) and Eq. \eqref{eq:XI}) due to the nonelectrostatic interaction. Accordingly, the profiles in (b) and (c) decay on the scale of $\kappa$ more rapidly than their counterparts in Figs. \ref{fig:DensitiesElementaryVectors1000}(d) and (f). For the remaining relevant parameters see Sec. \ref{Sec:ChoiceOfParameters}.\label{fig:DensitiesElementaryVectors0001}}
\end{figure}

After having discussed the effects of the boundary condition $\hat{\vec{h}} \neq 0$ via Figs. \ref{fig:density1000} and \ref{fig:DensitiesElementaryVectors1000},
the following second type of boundary condition is analyzed:
\begin{align}
	\hat{\vec{h}}(\vec{q_\|}) &= \vec{0},\nonumber \\
	\hat{\sigma}(\vec{q_\|}) = \sigma^{(0)},\text{ i.e., } &\sigma (\vec{r_\|})=\sigma^{(0)} ~\delta(\vec{r_\|}),
	\label{eq:BC0001Eq}
\end{align}
leading to 
\begin{equation}
	\vec{v}'(\vec{q_\|},0) = -\frac{\beta e \sigma^{(0)}}{\epsilon_0\epsilon}(0,0,0,1).
	\label{eq:BC0001Eq2}
\end{equation}
As before, the physical realization of this boundary condition is a $\delta$-like interaction, with the only difference residing in the type
of the basic interaction. Unlike in the previous case, here the interaction is of electrostatic character. Thus the situation corresponds to a $\delta$-like
negative charge distribution placed at the origin of the wall. Since the two ion types respond oppositely, the ion density profiles differ only in sign:
\begin{equation}
	\delta\hat{\rho}_2 = - \delta\hat{\rho}_3.
\end{equation}
This implies that the total ion density deviations vanish $\delta\widehat{\rho}_2 + \delta\widehat{\rho}_3 = 0$. Accordingly, also the 
density deviation for the solvent vanishes, $\delta\hat{\rho}_1 = 0.$
Figure \ref{fig:DensitiesElementaryVectors0001} shows the density profiles of the positive ions, which, up to the sign, are the same as the ones for the negative ions.
As stated above, for this boundary condition, there is no need to discuss the behavior of the solvent particles.

The three panels in Fig. \ref{fig:DensitiesElementaryVectors0001} are obtained similarly as the ones in Fig. \ref{fig:DensitiesElementaryVectors1000}. Figure \ref{fig:DensitiesElementaryVectors0001} (a)
shows the density profiles $\delta\widehat{\rho_2}(\vec{q_\|},z)$ as functions of the normal distance $z$ from the wall for three values of the lateral wave number $|\vec{q_\|}|$.
Figure \ref{fig:DensitiesElementaryVectors0001} (b) shows the same density profiles $\delta\widehat{\rho_2}(\vec{q_\|},z)$ but as functions of $|\vec{q_\|}|$ for three distances $z$ from the wall. 
Figure \ref{fig:DensitiesElementaryVectors0001} (c) displays the double Fourier transform $\delta\widehat{\widehat{\rho}}_2(\vec{q_\|},q_z)$.
Compared with the profiles in Fig. \ref{fig:DensitiesElementaryVectors1000} for the previously discussed boundary conditions, all present profiles differ significantly from them. Figure \ref{fig:DensitiesElementaryVectors0001} (a)
reveals a much larger decay length in $z$-direction, i.e., normal to the wall. Also in Fourier space the decay in lateral direction occurs much more rapidly, i.e., on much longer length scales in real space
than in the case of the nonelectrostatic wall-fluid interaction (c.f. Fig. \ref{fig:DensitiesElementaryVectors1000}).
This is also indicated by the much narrower peak in the double Fourier transform (see Fig. \ref{fig:DensitiesElementaryVectors0001} (c)). Furthermore, Fig. \ref{fig:DensitiesElementaryVectors0001} (a) shows a
variation of the decay length in normal direction as function of $|\vec{q}_\||$. In Fig. \ref{fig:DensitiesElementaryVectors0001} (b) one observes that the lateral wave numbers $|\vec{q_\|}|$ at
which the profiles $\delta\widehat{\rho_2} = -\delta\widehat{\rho_3}$ decay to half of the maximum values decrease upon increasing the distances $z$ from the wall, from which
one infers that the lateral decay length in real space increases upon increasing $z$.
The decay with respect to $|\vec{q}_\||$ is much faster than in the previous case (compare Fig. \ref{fig:DensitiesElementaryVectors1000} (d)),
indicating that in real space there is a slower decay in the lateral direction. Moreover, in Figs. \ref{fig:DensitiesElementaryVectors0001} (b) and \ref{fig:DensitiesElementaryVectors0001} (c) the functional form 
differs from the one shown in Figs. \ref{fig:density1000} and \ref{fig:DensitiesElementaryVectors1000}. These differences naturally occur due to the different
form of the boundary condition. Since in the case of the boundary condition studied above (see Figs. \ref{fig:density1000} and \ref{fig:DensitiesElementaryVectors1000}) the relevant interaction is nonelectrostatic,
the length scale dominating the decay is given by the corresponding short-ranged bulk correlation length $\xi$ (see Eq. \eqref{eq:XI}).
In contrast, for the system shown in Fig. \ref{fig:DensitiesElementaryVectors0001}, due to the electrostatic nature of the corresponding interaction, the dominating length scale is the Debye length
$1/\kappa$ (see Eq. \eqref{eq:DEBYE}), which is much larger than the correlation length $\xi$ due to the nonelectrostatic interaction, giving rise to the much slower decay
in Fig. \ref{fig:DensitiesElementaryVectors0001} (a) (on the scale of $1/\kappa$) and the much faster decay in Figs. \ref{fig:DensitiesElementaryVectors0001} (b) and (c) (on the scale of $\kappa$).

\subsection{Circular patch of interaction\label{Sec:ResultsPatch}}

\begin{figure*}
	\includegraphics{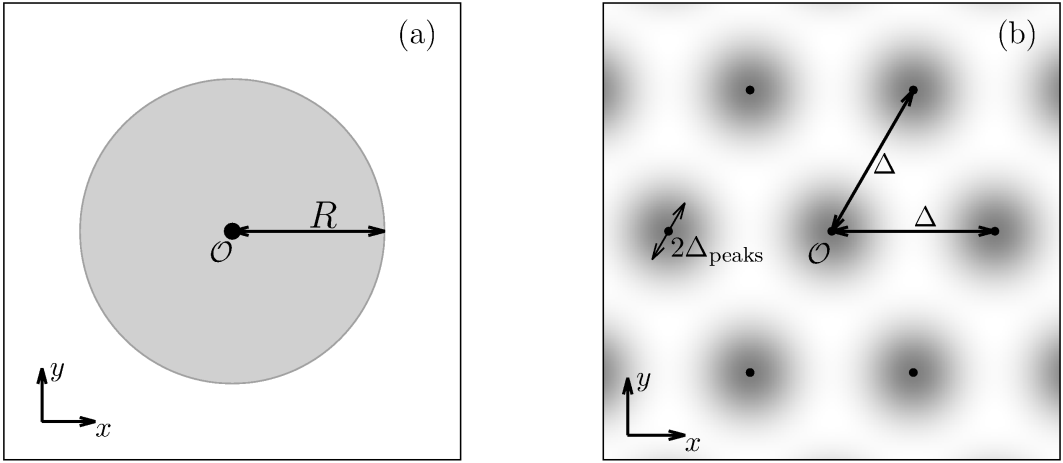}
	\caption{Physical configurations studied in Sec. \ref{Sec:ResultsPatch} (a) and in Sec. \ref{Sec:ResultsCrystal} (b). In Sec. \ref{Sec:ResultsPatch}, a two-dimensional circular patch of radius $R$ centered at the origin is studied (a). The dots in panel (b) correspond to the positions of the centers of the Gaussian interaction sites for the model used in Sec. \ref{Sec:ResultsCrystal}, which form a two-dimensional hexagonal lattice with lattice constant $\Delta$. The variance of the Gauss distributions is $\Delta^2_{\text{peaks}}$ (Eqs. \eqref{eq:Crystalh} and \eqref{eq:Crystalsigma}) . Our results are based on the choice $\Delta = 5 \Delta_{\text{peaks}}$ (see, c.f., Fig. \ref{fig:DensitiesCrystal}). \label{fig:InfoPatchAndCrystal}}
\end{figure*}

Having discussed in Sec. \ref{Sec:ResultsBasisVector} actually point-like interactions between the wall and the fluid, as the next step we now study the influence of interaction patterns
on the density deviations close to a wall upon broadening the spatial extent of the interaction area. To this end we analyze the influence of a two-dimensional circular interaction patch of radius $R$ 
centered at the origin (see Fig. \ref{fig:InfoPatchAndCrystal} (a)).

\begin{figure*}
	\includegraphics{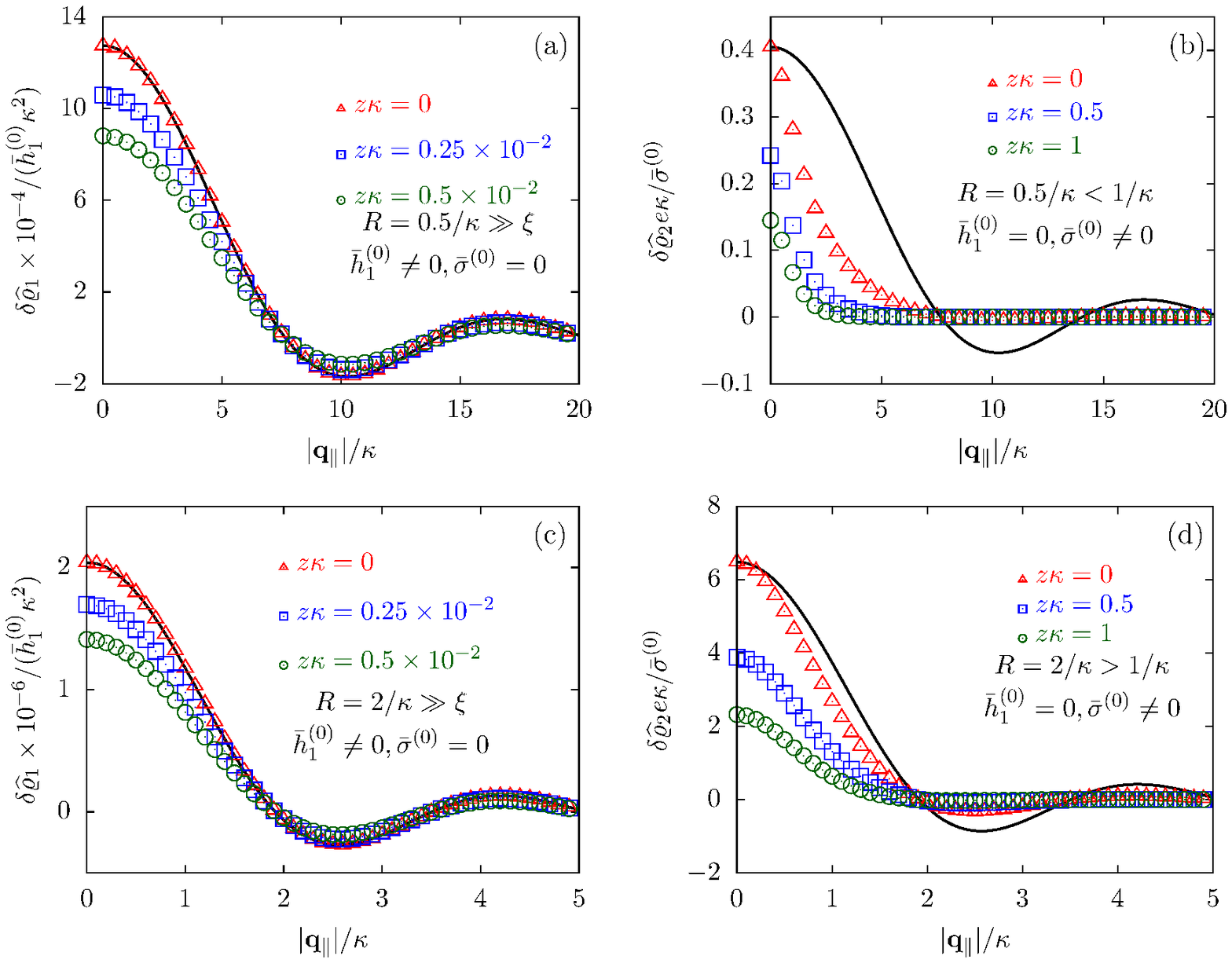}
	\caption{Density profiles of the solvent (panels (a) and (c)) and of the ions (panels (b) and (d)) as functions of the absolute value $|\vec{q_\|}|$ of the lateral wave vector for three normal distances $z$ from the wall. The boundary condition corresponds to a circular interaction patch centered at the origin with radius $R >0$. In the panels (a) and (b) the radius of the patch is $R = 0.5/\kappa$ whereas in the panels (c) and (d) the radius is $R = 2/\kappa$, where $1/\kappa$ is the Debye length (see Eq. \eqref{eq:DEBYE}). All considered patch radii are much larger than the bulk correlation length, $R\gg\xi$ (see Eq. \eqref{eq:XI}). In addition, there are different types of interaction. In panels (a) and (c) the interaction between the wall and the solvent particles is nonelectrostatic ($h_1(\vec{r_\|}) = \bar{h}_1^{(0)}\Theta(R-|\vec{r_\|}|), h_2(\vec{r_\|}) = h_3(\vec{r_\|}) = \sigma(\vec{r_\|}) = 0$, see Eq. \eqref{eq:Patch1000} as well as Figs. \ref{fig:density1000} and \ref{fig:DensitiesElementaryVectors1000}), whereas in panels (b) and (d) the patch contains a constant surface charge and therefore interacts with the ions only ($\vec{h}(\vec{r_\|}) = 0, \sigma(\vec{r_\|}) = \bar{\sigma}^{(0)}\Theta(R-|\vec{r_\|}|)$, see Eq. \eqref{eq:Patch0001} and Fig. \ref{fig:DensitiesElementaryVectors0001}). Besides the profiles, all panels show also the lateral Fourier transform of the boundary condition (Eqs. \eqref{eq:BCPatch1000} and \eqref{eq:BCPatch0001}) displayed as a black solid line. In the case of the interaction of the wall with the solvent ((a) and (c)), the decay of the profiles as function of $|\vec{q_\|}|$ is proportional to the lateral Fourier transform of the boundary condition, which implies, that the density deviations in real space closely follow the shape of the patch. However, in the case of a charged patch at the surface the density distribution of the ions reflects the competition between the length scale R of the radius of the patch and the Debye length $1/\kappa$. In the case of small patches ($R < 1/\kappa$, panel (b)), the Debye length dominates and therefore dictates the decay as function of $|\vec{q_\|}|$ without noticeable influence of the patch. In contrast, in the case of large patches ($R>1/\kappa$, panel (d)), in which the radius of the patch is the dominating length scale, the shape of the profiles follows the Fourier transform of the charge distribution at the wall, i.e., the patch of radius $R$. For the remaining relevant parameters see Sec. \ref{Sec:ChoiceOfParameters}.\label{fig:DensitiesPatch}}
\end{figure*}

Due to the radial symmetry, the spatial structures in Fourier space depend only on the absolute value $|\vec{q_\|}|$ of the lateral wave vector $\vec{q_\|}$.
Figure \ref{fig:DensitiesPatch} discusses four distinct configurations.

Alluding to the insights gained in the previous section, Figs. \ref{fig:DensitiesPatch} (a) and (c) correspond to a homogeneous circular patch of radius $R$, which interacts with the solvent only,
similar to Figs. \ref{fig:density1000} and \ref{fig:DensitiesElementaryVectors1000}. This amounts to the boundary condition (see Eq. \eqref{eq:V_BC})
\begin{align}
	h_1(\vec{r_\|}) &= \bar{h}_1^{(0)} ~ \Theta(R-|\vec{r}|), \nonumber\\
	h_2(\vec{r_\|}) &= h_3(\vec{r_\|}) = \sigma(\vec{r_\|}) = 0
	\label{eq:Patch1000}
\end{align}
leading to
\begin{equation}
	\vec{v}'(\vec{q_\|},0) =  - 2\pi R^2 \frac{\bar{h}_1^{(0)}}{b}\frac{J_1(|\vec{q_\|}|R)}{|\vec{q_\|}|R}~(1,0,0,0),
	\label{eq:BCPatch1000}
\end{equation}
where the two-dimensional Fourier transform of the Heaviside function $\Theta(R - |\vec{r}|)$ is given by
\begin{equation}
	\int_{\mathbb{R}^2} \mathrm{d}^2 r_\|~\Theta(R - |\vec{r_\|}|) \exp(-i\vec{q_\|}\cdot\vec{r_\|}) = 2\pi R^2 \frac{J_1(|\vec{q_\|}|R)}{|\vec{q_\|}|R}.
\end{equation}
In contrast, Figs. \ref{fig:DensitiesPatch} (b) and (d) refer to a charged circular patch of radius $R$, similar to Fig. \ref{fig:DensitiesElementaryVectors0001}:
\begin{equation}
	\vec{h}(\vec{r_\|}) = 0,~ \sigma(\vec{r_\|}) = \bar{\sigma}^{(0)} \Theta(R-|\vec{r_\|}|),
	\label{eq:Patch0001}
\end{equation}
leading to the boundary condition
\begin{equation}
	\vec{v}'(\vec{q_\|},0) =  -2\pi R^2 \frac{\beta e\bar{\sigma}^{(0)}}{\epsilon_0\epsilon}\frac{J_1(|\vec{q_\|}|R)}{|\vec{q_\|}|R}~(0,0,0,1).
	\label{eq:BCPatch0001}
\end{equation} 
Figures \ref{fig:DensitiesPatch} (a) and (b) correspond to the patch size $R = 0.5/\kappa$ whereas Figs. \ref{fig:DensitiesPatch} (c) and (d) correspond to $R=2/\kappa$.
In all four panels the black line is given by $A J_1(|\vec{q_\|}|R)/(|\vec{q_\|}|R)$ with $A$ chosen such that the first maximum of the data for $z\kappa = 0$ is reproduced.
In Figs. \ref{fig:DensitiesPatch}(a) and (c) only the solvent density profiles are shown, because, due to linearity, Sec. \ref{Sec:ResultsBasisVector} indicates, that the ion profiles are
proportional to the one of the solvent. Figures \ref{fig:DensitiesPatch} (a) and (c) clearly show, that the density deviations are proportional to the Fourier transform $\vec{v}'(\vec{q_\|},0)$ of
the boundary condition (Eq. \eqref{eq:BCPatch1000}, solid black line). This implies that for increasing lateral distances from the center of the patch the decay of the profiles in real space is dominated by
the length scale set by the radius $R$ of the patch. This trend holds for both patch sizes. However, as expected, the amplitudes of the density deviations increase for the
larger patch size (note the different scales). In contrast, in Figs. \ref{fig:DensitiesPatch} (b) and (d), where the density profiles of the positive ions are shown and where the profiles for the solvent are
omitted for the same reasons as explained in Sec. \ref{Sec:ResultsBasisVector}, the profiles do not follow the Fourier transform of the boundary conditions (solid black line). This is
particularly pronounced in Fig. \ref{fig:DensitiesPatch} (b), i.e., for the smaller patch size. In this case the decay as function of $|\vec{q_\|}|$ is faster than the Fourier transform of the boundary
condition, which implies that the profiles decay on a length scale larger than that of the radius $R$ of the patch and also the shape of the decay differs from that of the expression $J_1(|\vec{q_\|}|R)/(|\vec{q_\|}|R)$.
This behavior can be understood in terms of two distinct dominating length scales. In Figs. \ref{fig:DensitiesPatch} (b) and (d), where the effect of electrostatic interactions
are shown, the dominating length scale is the Debye length $1/\kappa$ in contrast to the much smaller correlation length $\xi$ ($\xi \approx 1.3 \times 10^{-2}~\kappa^{-1}$) induced by the nonelectrostatic
interactions characterizing Figs. \ref{fig:DensitiesPatch} (a) and (c). Since in Fig. \ref{fig:DensitiesPatch} (b) the radius $R$ of the patch is only half the Debye length $1/\kappa$,
the dominating length scale is the Debye length $1/\kappa$, so that the density deviations decay in real space on a length scale which is larger than the patch radius $R$.
Also, since the profile in Fourier space is not proportional to the Fourier transform of the boundary condition, one can conclude that the shape of the patch has no
significant influence on the decay behavior. The competition of the length scales $\xi$, $1/\kappa$, and $R$ is also borne out in Fig. \ref{fig:DensitiesPatch} (d),
where the patch size $R$ is twice as large as the Debye length $1/\kappa$. This case is much more similar to the ones in Figs.~\ref{fig:DensitiesPatch} (a) and (c), because the dominating length scale
is set by the radius $R$, and consequently the profiles follow rather closely the shape (solid black line) dictated by the interaction patch. However, the influence of the smaller Debye length
scale is still visible, which is the reason for the deviations from the Fourier transform of the boundary condition (solid black line).
In conclusion, as already seen in Sec. \ref{Sec:ResultsBasisVector}, the largest length scale sets the decay behavior of the density deviations. In the present
case of non-vanishing sizes of the interaction areas, the largest length scale dictates not only the range
but also the shape of the density deviations.

\subsection{Periodic distribution of interaction sites \label{Sec:ResultsCrystal}}

After having discussed the density profiles in the presence of spatially localized, single interaction sites in Secs. \ref{Sec:ResultsBasisVector} and \ref{Sec:ResultsPatch}, here we study the influence of 
interaction sites forming a regular hexagonal lattice:
\begin{equation}
	\vec{r}_{\text{peaks}} = (\alpha \Delta + \frac{\beta}{2}\Delta, \frac{\sqrt{3}}{2}\beta\Delta),\quad\alpha,\beta\in\mathbb{Z},
\end{equation}
see Fig. \ref{fig:InfoPatchAndCrystal} (b); the distance between nearest neighbor sites is denoted as $\Delta$.

\begin{figure*}
	\includegraphics{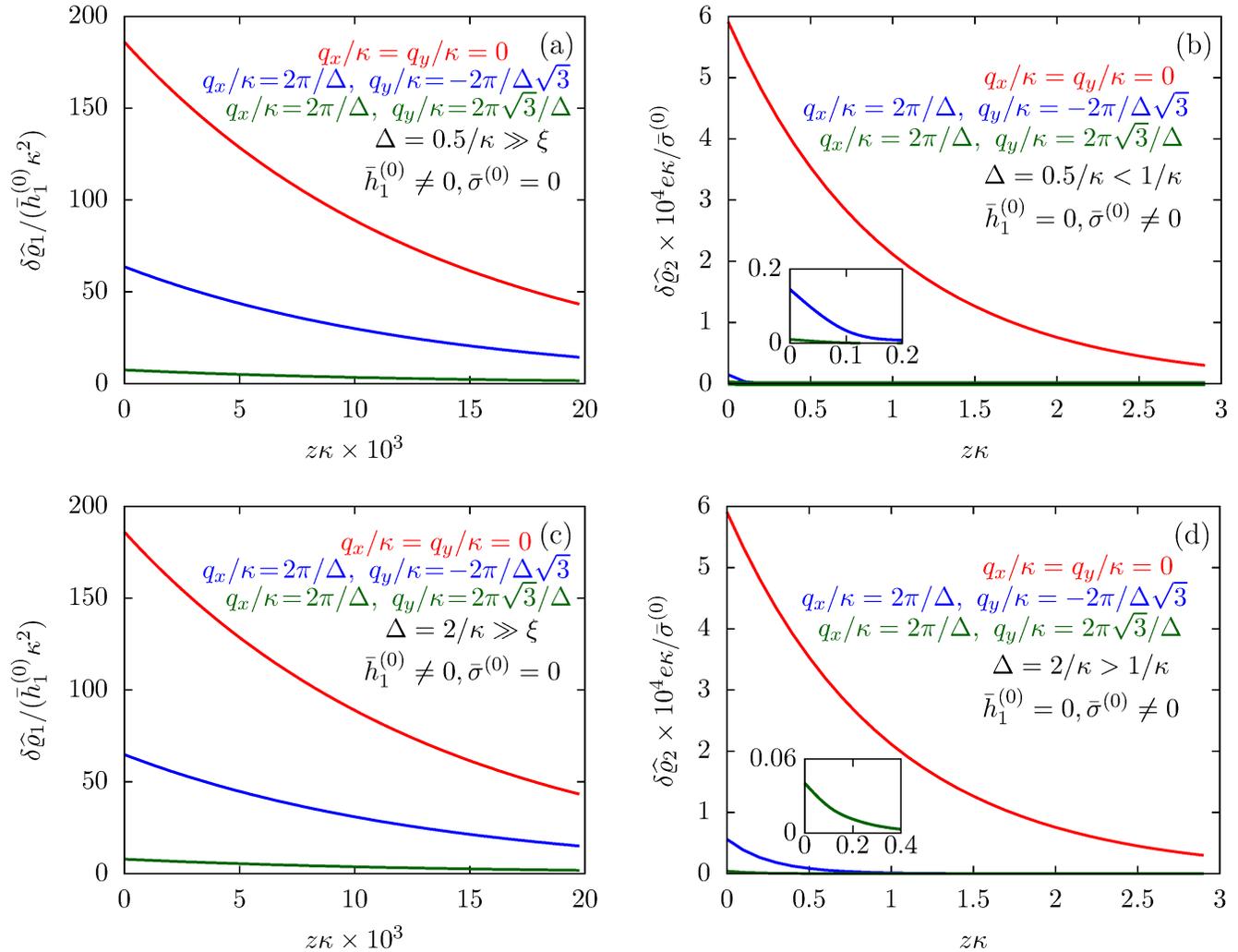}
	\caption{Density profiles of the solvent (panels (a) and (c)) and of the ions (panels (b) and (d)) for three lateral wave vectors $\vec{q_\|} = (q_x,q_y)$ as functions of the normal distance $z$ from the wall in units of the Debye length $1/\kappa$. The boundary condition corresponds to a hexagonal lattice of interaction sites with a Gaussian charge distribution characterized by a standard deviation $\Delta_{\text{peaks}} = \Delta/5$. The lattice constant is denoted as $\Delta$ (see Fig. \ref{fig:InfoPatchAndCrystal}). In panels (a) and (b) the lattice constant and the variance are $\Delta = 0.5/\kappa$ and $\Delta_{\text{peaks}} = 0.1/\kappa$, respectively, whereas in panels (c) and (d) the lattice constant and the variance are $\Delta = 2/\kappa$ and $\Delta_{\text{peaks}} = 0.4/\kappa$, respectively, with the Debye length $1/\kappa$ (see Eq. \eqref{eq:DEBYE}). Panels (a) and (c) correspond to systems with a nonelectrostatic interaction between the wall and the solvent particles (see Eq. \eqref{eq:BC1Crystal}), whereas panels (b) and (d) correspond to systems with electrostatic interaction sites between wall and ions (see Eq. \eqref{eq:BC4Crystal}). The insets in (b) and (d) show a magnified version of the respective profiles in the main plot. In all cases, the profiles decay exponentially upon increasing the normal distance $z$ from the wall. However, the decay length differs significantly between the two aforementioned types of interactions. In the case of the nonelectrostatic interaction, the decay length is set by the bulk correlation length $\xi$ (see Eq. \eqref{eq:XI}) of the fluid, whereas in the case of the electrostatic interaction it is set by the much larger Debye length $1/\kappa$. This difference in the decay lengths, both in lateral and in normal direction, which leads to a much faster lateral decay in the case of the electrostatic interactions, is also responsible for the decreasing amplitude of the ion profiles for increased wave vectors (panel (b) and (d)). Another significant difference between the two interaction types is the variation of the decay length as function of the lateral wave vectors. In panels (a) and (c) all profiles decay exponentially on the same decay length $\xi$, whereas in panels (b) and (d) the decay length depends significantly on the wave vectors. This effect follows from the dependence of the eigenvalues on $|\vec{q_\|}|$ as discussed in Eq. \eqref{eq:EWH}, corresponding to a lateral decay proportional to $\exp(-\sqrt{\kappa^2 + |\vec{q_\|}|^2}z)$. In principle this occurs for both types of interactions. However only in the case of the electrostatic interactions it is relevant, which again is due to the difference between the dominating length scales. For the remaining relevant parameters see Sec. \ref{Sec:ChoiceOfParameters}.\label{fig:DensitiesCrystal}}
\end{figure*}

Distinct from the previous examples in Secs. \ref{Sec:ResultsBasisVector} and \ref{Sec:ResultsPatch}, the interaction strength around the individual interaction sites $\vec{r}_{\text{peaks}}$ is taken to form
a Gaussian distribution, providing either a nonelectrostatic or an electrostatic interaction with equal amplitudes for all interaction sites:
\begin{align}
	h_i(\vec{r}) &= \bar{h}_i^{(0)} ~ \sum_{\text{peaks}} \exp\left(-\frac{(\vec{r} - \vec{r}_{\text{peaks}})^2}{2 \Delta^2_{\text{peaks}}}\right), \quad i = 1,2,3,
	\label{eq:Crystalh}
\end{align}
and
\begin{align}	
	\sigma(\vec{r}) &= \bar{\sigma}^{(0)} ~ \sum_{\text{peaks}} \exp\left(-\frac{(\vec{r} - \vec{r}_{\text{peaks}})^2}{2 \Delta^2_{\text{peaks}}}\right),
	\label{eq:Crystalsigma}
\end{align}
respectively, where $\Delta^2_{\text{peaks}}$ is the variance of the Gaussian interaction. Lateral Fourier transformation leads to the corresponding boundary condition $\vec{v}'$ (see Eq. \eqref{eq:V_BC}) with
\begin{align}
	v'_i (\vec{q_\|}) =& -\frac{\bar{h}_i^{(0)}}{b} (2 \pi \Delta^2_{\text{peaks}}) \exp\left(-\frac{\vec{q_\|}^2\Delta^2_{\text{peaks}}}{2}\right) |\mathcal{C}_{\mathcal{G}}|\times\nonumber\\
	&\sum_{\vec{G}\in\mathcal{G}}\delta(\vec{q_\|} - \vec{G}), \quad i = 1,2,3,
\end{align}
and
\begin{align}
	v'_4 (\vec{q_\|}) =& -\frac{\beta e \bar{\sigma}^{(0)}}{\epsilon_0\epsilon} (2 \pi \Delta^2_{\text{peaks}}) \exp\left(-\frac{\vec{q_\|}^2\Delta^2_{\text{peaks}}}{2}\right) |\mathcal{C}_{\mathcal{G}}|\times\nonumber\\
			   &\sum_{\vec{G}\in\mathcal{G}}\delta(\vec{q_\|} - \vec{G}),
\end{align}
where $|\mathcal{C}_{\mathcal{G}}| = (16\pi^2/(3\Delta^2)) \sin(\SI{60}{\degree})$ is the size of an elementary cell of the corresponding two-dimensional reciprocal lattice $\mathcal{G}$. 
Using this boundary condition, we have studied four different systems, as shown in Fig. \ref{fig:DensitiesCrystal}.

The four panels are arranged as in Fig. \ref{fig:DensitiesPatch}, with the boundary conditions corresponding to an interaction between the wall and solvent
only in Figs. \ref{fig:DensitiesCrystal} (a) and (c),
\begin{align}
	\vec{v}'(\vec{q_\|}) =&  -\frac{\bar{h}_1^{(0)}}{b}(2 \pi \Delta^2_{\text{peaks}}) \exp\left(-\frac{\vec{q_\|}^2\Delta^2_{\text{peaks}}}{2}\right) |\mathcal{C}_{\mathcal{G}}|\times\nonumber\\
			      &\sum_{\vec{G}\in\mathcal{G}}\delta(\vec{q_\|} - \vec{G})~(1,0,0,0),
	\label{eq:BC1Crystal}
\end{align}
and in Figs. \ref{fig:DensitiesCrystal} (b) and (d) boundary conditions corresponding to an interaction between the wall and ions only, i.e., due to a hexagonal lattice of interaction sites with Gaussian
intrinsic charge distribution:
\begin{align}
	\vec{v}'(\vec{q_\|}) =& \frac{\beta e \bar{\sigma}^{(0)}}{\epsilon_0 \epsilon} (2 \pi \Delta^2_{\text{peaks}}) \exp\left(-\frac{\vec{q_\|}^2\Delta^2_{\text{peaks}}}{2}\right) |\mathcal{C}_{\mathcal{G}}|\times\nonumber\\
			      &\sum_{\vec{G}\in\mathcal{G}}\delta(\vec{q_\|} - \vec{G})~(0,0,0,1).
	\label{eq:BC4Crystal}
\end{align}
For the same reason as stated in the context of Fig. \ref{fig:DensitiesPatch}, in the former case (Figs. \ref{fig:DensitiesCrystal} (a) and (c)) only the deviations of the solvent density and in the latter case 
(Figs. \ref{fig:DensitiesCrystal} (b) and (d)) only the deviations of the ion densities are shown. Figures \ref{fig:DensitiesCrystal} (a) and (b) correspond to the lattice constant $\Delta = 0.5/\kappa$
whereas Figs. \ref{fig:DensitiesCrystal} (c) and (d) correspond to $\Delta = 2/\kappa$. The variance $\Delta^2_{\text{peaks}}$ of the peaks is
taken as $\Delta_{\text{peaks}} = \Delta/5$, so that $\Delta_{\text{peaks}} = 0.1/\kappa$ in Figs. \ref{fig:DensitiesCrystal} (a) and (b) and $\Delta_{\text{peaks}} = 0.4/\kappa$
in Figs. \ref{fig:DensitiesCrystal} (c) and (d).
Figures \ref{fig:DensitiesCrystal} (a) and (c) tell that, although different values for $\vec{q_\|}$ change the amplitude of the profiles in all cases, the solvent profile decays 
exponentially, upon increasing the normal distance $z$ from the wall, on the scale of the bulk correlation length $\xi$. This holds for both values of the lattice constant $\Delta$. However, the
amplitude of the density deviations is slightly increased for the larger lattice constant $\Delta$, which is in line with the also increased variance $\Delta^2_{\text{peak}}$ of the interaction sites.
In contrast to these findings, Figs. \ref{fig:DensitiesCrystal} (b) and (d) reveal a different picture. In these panels, one still finds an exponential decay of the profiles upon increasing
the normal distance $z$. However, the profiles decay on a much larger length scale than the ones in Figs. \ref{fig:DensitiesCrystal} (a) and (c). Moreover, not only the amplitude
but also the decay length changes significantly for different values of $\vec{q_\|}$. This was already encountered in Fig. \ref{fig:DensitiesElementaryVectors0001},
where the decay length depends on the value of $|\vec{q_\|}|$.  This variation of the decay lengths can be inferred from Eq. \eqref{eq:V}, which shows that the
eigenvalues and thus the decay length depends on $k = |\vec{q_\|}|$. The variation of the decay length can be expressed in terms of the Debye length $1/\kappa$, which determines the length scale in case of
$|\vec{q_\|}| = 0$. From Eq. \eqref{eq:EWH} one finds, that the decay as function of $z$ is proportional to $\exp(-\sqrt{\kappa^2 + |\vec{q_\|}|^2}z)$. The large differences in
the amplitudes of the various profiles in Figs. \ref{fig:DensitiesCrystal} (b) and (d), as well as the pronounced increase of the decay length in comparison to Figs. \ref{fig:DensitiesCrystal} (a) and (c) can be 
understood in terms of the differences between the dominating length scale. Analogous to the previous sections, for the systems shown in Figs. \ref{fig:DensitiesCrystal} (a) and (c),
the dominating length scale in lateral direction is the length scale set by the boundary conditions and the bulk correlation length $\xi$ which characterizes the decay of the solvent density in normal direction.
However, for the systems shown in Figs. \ref{fig:DensitiesCrystal} (b) and (d), the relevant inherent length scale of the fluid is the Debye length $1/\kappa$, which is significantly larger than the bulk correlation
length $\xi$ and thus causes the increase in the length scale of the decay, both in lateral and in normal direction.

\section{Conclusions and summary\label{Sec:ConclusionsSummary}}

In the present analysis the influence of a chemically or electrostatically structured surface on an adjacent fluid has been studied and described in terms of the density profiles of the fluid components. The fluid, which comprises a single
solvent species and a single univalent salt far away from bulk and wetting phase transitions, has been investigated within classical density functional theory \cite{Evans1979, Evans1990, Evans1992}. 
Within this model four examples of heterogeneous walls have been studied. First, single isolated interaction sites are discussed, which interact either nonelectrostatically (between the wall and solvent particles)
or electrostatically (between the wall and ions) (see Secs. \ref{Sec:ResultsBasisVector} and \ref{Sec:ResultsPatch}). In the case of a $\delta$-like nonelectrostatic
interaction, the solvent density increases around the interaction site and decays exponentially on the length scale of the bulk correlation length $\xi$.
The deviations of the ion number densities from their bulk values are proportional to that of the number density of the solvent (see Figs. \ref{fig:density1000} and \ref{fig:DensitiesElementaryVectors1000}). For a $\delta$-like electrostatic
interaction, within the present model, the solvent does not respond at all, because the deviations induced by the two ion types even out due to symmetries, whereas the density deviations of the ionic particles
again decay exponentially. However, the length scale of the latter decay is significantly increased as compared to the former case, because the dominating scale in this case is
the Debye length $1/\kappa \gg \xi$ (see Fig. \ref{fig:DensitiesElementaryVectors0001}).
The introduction of another length scale by studying interaction sites of non-vanishing extent (see Sec. \ref{Sec:ResultsPatch}) shows, that the resulting density profiles strongly depend on the dominant length scale (see Fig.
\ref{fig:DensitiesPatch}). If a bulk length scale (bulk correlation length $\xi$ or Debye length $1/\kappa$) dominates, the profiles resemble the ones for $\delta$-like interactions. However, if a
length scale set by a boundary condition at the wall dominates or is similar to the dominating length scale in the bulk, the decay of the density deviations increasingly reflects the boundary conditions.
Finally, the examination of multiple interaction sites, arranged as a regular hexagonal lattice (see Fig. \ref{fig:InfoPatchAndCrystal} (b)), shows, that the size of the interaction sites
and the distance between them influence the amplitude and thus the importance of density deviations for large values of the lateral wave number $|\vec{q}_\||$ (see Fig. \ref{fig:DensitiesCrystal}).

In summary, the present study provides a flexible framework to determine the influence of various surface inhomogeneities on the density profiles of a fluid in contact with that substrate. The
resulting profiles are found to be sensitive to the type of interaction as well as to the size and the distribution of the interaction sites. 

This framework is considered as a starting point for extensions into various directions, aiming for the analysis of more sophisticated and realistic models. First, the model used here to describe the fluid is a
very simple one, chosen to lay a foundation for further research and to introduce the approach as such. Concerning future work, more realistic descriptions of the fluid and more elaborate density functional descriptions
could be used. For instance, the present restriction to low ionic strengths and equal particle sizes can be removed along the lines of Ref.~\cite{Podgornik2016}. Second, for the systems studied here, the fluid is thermodynamically far from any bulk or wetting phase transitions. This is solely done for the sake of simplicity. In future studies 
of more realistic systems, taking into account the occurrence of phase transitions and their influence on the systems is expected to be rewarding. Third, this study is restricted to linear response theory.
Whereas this allows for a broad overview of structure formation in terms of superpositions of only a few elementary patterns, the occurrence of nonlinear structure formation phenomena
requires approaches beyond linear response theory. Finally, studying the influence of
disordered surface structures within the present framework appears to be very promising.

\appendix

\section{Eigenvectors and eigenvalues of $\mat{H}(k)$\label{app:EVEW}}

According to the structure of the matrix $\mat{M}$ (Eq. \eqref{eq:local}), with entries given by Eq.~(\ref{eq:Mij}), and of the vector
$\vec{Z}=(0,1,-1)$, from Eqs.~(\ref{eq:MatELG1}) and (\ref{eq:MatELG2}) one infers that the matrix $\mat{H}(k)$ has the form
\begin{align}
	\mat{H}(k) = 
	\left(\begin{array}{cccc}
			s & u  & u   & 0   \\
			u & t  & u   & iv  \\
			u & u  & t   & -iv \\
			0 & iv & -iv & 0   
	\end{array}\right) + \text{diag}(k^2)
	\label{eq:structH}
\end{align}
with $s,t,u,v\in\mathbb{R}$ and $k = |\vec{q_\|}|,s,t>0$.
It can be readily verified that the four vectors
\begin{align}
	\vec{\Lambda}_1(k) &:= (2u, \lambda_1(k)-s, \lambda_1(k)-s, 0), \notag\\
	\vec{\Lambda}_2(k) &:= (\lambda_2(k)-t-u, u, u, 0),             \notag\\
	\vec{\Lambda}_3(k) &:= (0, \lambda_3(k), -\lambda_3(k), 2iv),    \notag\\
	\vec{\Lambda}_4(k) &:= (0, iv, -iv, \lambda_4(k)-t+u)
	\label{eq:EVH}
\end{align}
with $\vec{\Lambda}_i \in \mathbb{C}^4 \text{ for } i=1,\dots,4$, form a \emph{nonorthogonal} basis of eigenvectors of the matrix
$\mat{H}(k)$ in Eq.~(\ref{eq:structH}) with the respective real eigenvalues
\begin{align}
	\lambda_1(k) &= \frac{1}{2}\left(s+t+u + \sqrt{(s-t-u)^2 + 8u^2}\right) + k^2, \notag\\
	\lambda_2(k) &= \frac{1}{2}\left(s+t+u - \sqrt{(s-t-u)^2 + 8u^2}\right) + k^2, \notag\\
	\lambda_3(k) &= \frac{1}{2}\left(t-u + \sqrt{(t-u)^2 - 8v^2}\right) + k^2,     \notag\\
	\lambda_4(k) &= \frac{1}{2}\left(t-u - \sqrt{(t-u)^2 - 8v^2}\right) + k^2.
	\label{eq:EWH}
\end{align}

The expressions for $s,t,u, \text{ and } v$ can be obtained from the bulk quantities mentioned in Sec.~\ref{Sec:ChoiceOfParameters} and
take on the forms (see Eqs. \eqref{eq:b} and \eqref{eq:Mij}) 
\begin{align}
	s &= \frac{M_{11}}{b}, \\
	t &= \frac{M_{22}}{b} = \frac{M_{33}}{b}, \\
	u &= \frac{M_{12}}{b} = \frac{M_{13}}{b} = \frac{M_{23}}{b}, \\
	v &= -\sqrt{\frac{4\pi l_B}{\epsilon b}}.
\end{align}


\begin{thebibliography}{00}

	\bibitem{Bagotsky2006}
		V.S.\ Bagotsky,
		\textit{Fundamentals of Electrochemistry} 
		(Wiley, Hoboken, 2006).

	\bibitem{Schmickler2010}
		W.\ Schmickler and E.\ Santos,
		\textit{Interfacial Electrochemistry} 
		(Springer, Berlin, 2010).

	\bibitem{Dietrich1988}
		S.\ Dietrich,
		\textit{Wetting phenomena},
		in \textit{Phase Transitions and Critical Phenomena}, Vol.\ 12,
		edited by C.\ Domb and J.L.\ Lebowitz
		(Academic, London, 1988), p.\ 1.

	\bibitem{Schick1990}
		M.\ Schick,
		\textit{Introduction to wetting phenomena},
		in \textit{Les Houches, Session XLVIII, 1988 --- Liquides aux interfaces / Liquids at 
		interfaces},
		edited by J.\ Charvolin, J.F.\ Joanny, and J.\ Zinn-Justin
		(North-Holland, Amsterdam, 1990), p.\ 415.

	\bibitem{Wen2017}
		M.\ Wen and K.\ Du\v{s}ek (eds.),
		\textit{Protective Coatings}
		(Springer, Cham, 2017).

	\bibitem{Vogel2012}
		N.\ Vogel,
		\textit{Surface Patterning with Colloidal Monolayers}
		(Springer, Berlin, 2012).

	\bibitem{Nee2015}
		A.Y.C.\ Nee,
		\textit{Handbook of Manufacturing Engineering and Technology}
		(Springer, London, 2015).

	\bibitem{Russel1989} 
		W.\ Russel, D.\ Saville, and W.\ Schowalter, 
		\textit{Colloidal Dispersions} 
		(Cambridge University, Cambridge, 1989).

	\bibitem{Hunter2001}
		R.J.\ Hunter,
		\textit{Foundations of Colloid Science}
		(Oxford University, Oxford, 2001).

	\bibitem{Lin2011}
		B.\ Lin (ed.), 
		\textit{Microfluidics}
		(Springer, Berlin, 2011).

	\bibitem{GalindoRosales2018}
		F.J.\ Galindo-Rosales,
		\textit{Complex Fluids and Rheometry in Microfluidics},
		in \textit{Complex Fluid-Flows in Microfluidics},
		edited by F.J.\ Galindo-Rosales
		(Springer, Cham, 2018), p.\ 1.

	\bibitem{Andelman1991}
		D.\ Andelman,
		\textit{On the Adsorption of Polymer Solutions on Random Surfaces: The Annealed Case},
		Macromolecules \textbf{24}, 6040 (1991).

	\bibitem{Chen2005}
		W.\ Chen, S.\ Tan, T.-K.\ Ng, W.T.\ Ford, and P.\ Tong, 
		\textit{Long-ranged attraction between charged polystyrene spheres at aqueous interfaces},
		Phys.\ Rev.\ Lett.\ \textbf{95}, 218301 (2005).

	\bibitem{Chen2006}
		W.\ Chen, S.\ Tan, S.\ Huang, T.-K.\ Ng, W.T.\ Ford, and P.\ Tong,
		\textit{Measured long-ranged attractive interaction between charged polystyrene latex spheres
		at a water-air interface},
		Phys.\ Rev.\ E \textbf{74}, 021406 (2006).

	\bibitem{Chen2009}
		W.\ Chen, S.\ Tan, Y.\ Zhou, T.-K.\ Ng, W.T.\ Ford, and P.\ Tong,
		\textit{Attraction between weakly charged silica spheres at a water-air interface induced by
		surface-charge heterogeneity},
		Phys.\ Rev.\ E \textbf{79}, 041403 (2009).

	\bibitem{Naji2010}
		A.\ Naji, D.S.\ Dean, J.\ Sarabadani, R.R.\ Horgan, and R.\ Podgornik,
		\textit{Fluctuation-induced interaction between randomly charged dielectrics},
		Phys.\ Rev.\ Lett.\ \textbf{104}, 060601 (2010).

	\bibitem{Ben-Yaakov2013}
		D.\ Ben-Yaakov, D.\ Andelman, and H.\ Diamant, 
		\textit{Interaction between heterogeneously charged surfaces: surface patches and charge
		modulation},
		Phys.\ Rev.\ E \textbf{87}, 022402 (2013).

	\bibitem{Naji2014}
		A.\ Naji, M.\ Ghodrat, H.\ Komaie-Moghaddam, and R.\ Podgornik,
		\textit{Asymmetric Coulomb fluids at randomly charged dielectric interfaces: anti-fragility,
		overcharging and charge inversion},
		J.\ Chem.\ Phys.\ \textbf{141}, 174704 (2014).

	\bibitem{Bakhshandeh2015}
		A.\ Bakhshandeh, A.P.\ dos Santos, A.\ Diehl, and Y.\ Levin, 
		\textit{Interaction between random heterogeneously charged surfaces in an electrolyte
		solution},
		J.\ Chem.\ Phys.\ \textbf{142}, 194707 (2015).

	\bibitem{Ghodrat2015a}
		M.\ Ghodrat, A.\ Naji, H.\ Komale-Moghaddam, and R.\ Podgornik, 
		\textit{Ion-mediated interactions between net-neutral slabs: weak and strong disorder 
		effects},
		J.\ Chem.\ Phys.\ \textbf{143}, 234701 (2015).

	\bibitem{Ghodrat2015b}
		M.\ Ghodrat, A.\ Naji, H.\ Komaie-Moghaddama, and R.\ Podgornik,
		\textit{Strong coupling electrostatics for randomly charged surfaces: antifragility and 
		effective interactions},
		Soft Matter \textbf{11}, 3441 (2015).

	\bibitem{Adar2016}
		R.M.\ Adar and D.\ Andelman,
		\textit{Electrostatic attraction between overall neutral surfaces},
		Phys.\ Rev.\ E \textbf{94}, 022803 (2016).

	\bibitem{Adar2017a}
		R.M.\ Adar and D.\ Andelman,
		\textit{Osmotic pressure between arbitrarily charged surfaces: a revisited approach},
		arXiv:1709.02114 (2017).

	\bibitem{Adar2017b}
		R.M.\ Adar, D.\ Andelman, and H.\ Diamant,
		\textit{Electrostatics of patchy surfaces},
		Adv.\ Colloid Interface Sci.\ \textbf{247}, 198 (2017).

	\bibitem{Ghosal2017}
		S.\ Ghosal and J.D.\ Sherwood,
		\textit{Screened Coulomb interactions with non-uniform surface charge},
		Proc.\ Roy.\ Soc.\ A \textbf{473}, 20160906 (2017).

	\bibitem{Zhou2017}
		S.\ Zhou,
		\textit{Effective electrostatic interactions between two overall neutral surfaces with
		quenched charge heterogeneity over atomic length scale},
		J.\ Stat.\ Phys.\ \textbf{169}, 1019 (2017).

	\bibitem{Onuki2004}
		A.\ Onuki and H.\ Kitamura,
		\textit{Solvation effects in near-critical binary mixtures},
		J.\ Chem.\ Phys.\ \textbf{121}, 3143 (2004).

	\bibitem{Bier2012a}
		M.\ Bier, A.\ Gambassi, and S.\ Dietrich,
		\textit{Local theory for ions in binary liquid mixtures},
		J.\ Chem.\ Phys.\ \textbf{137}, 034504 (2012). 

	\bibitem{Bier2012b}
		M.\ Bier and L. Harnau,
		\textit{The structure of fluids with impurities},
		Z.\ Phys.\ Chem.\ \textbf{226}, 807 (2012).

	\bibitem{Evans1979}
		R.\ Evans,
		\textit{The nature of the liquid-vapour interface and other topics in the statistical
		mechanics of non-uniform, classical fluids}, 
		Adv.\ Phys.\ \textbf{28}, 143 (1979).

	\bibitem{Evans1990} 
		R.\ Evans, 
		\textit{Microscopic theories of simple fluids and their interfaces},
		in \textit{Les Houches, Session XLVIII, 1988 --- Liquides aux interfaces / Liquids at
		interfaces}, 
		edited by J.\ Charvolin, J.F.\ Joanny, and J.\ Zinn-Justin 
		(North-Holland, Amsterdam, 1990), p. 1.

	\bibitem{Evans1992}
		R.\ Evans, 
		\textit{Density functionals in the theory of nonuniform fluids},
		in \textit{Fundamentals of inhomogeneous fluids}, 
		edited by D.\ Henderson
		(Marcel Dekker, New York, 1992), p. 85.

	\bibitem{Cahn1958}
		J.W.\ Cahn and J.E.\ Hilliard,
		\textit{Free energy of a nonuniform system. I. Interfacial free energy},
		J.\ Chem.\ Phys.\ \textbf{28}, 258 (1958).

	\bibitem{Bocquet2002}
		L.\ Bocquet, E.\ Trizac, and M.\ Aubouy,
		\textit{Effective charge saturation in colloidal suspension},
		J.\ Chem.\ Phys.\ \textbf{117}, 8138 (2002).

	\bibitem{Hansen1986}
		H.\ P.\ Hansen and I.\ R.\ McDonald,
		\textit{Theory of simple liquids, 2nd ed.}
		(Academic, San Diego, 1986).

	\bibitem{Naik2000}
		S.\ J.\ Suresh and V.\ M.\ Naik,
		\textit{Hydrogen bond thermodynamic properties of water from dielectric constant data},
		J.\ Chem.\ Phys.\ \textbf{113}, 9727 (2000).

	\bibitem{Lide1998}
		D.R.\ Lide,
		\textit{Handbook of Chemistry and Physics, 79th ed.}
		(CRC, Boca Raton, 1998). 

	\bibitem{harvey2004correlation}
		A.\ H.\ Harvea and E.\ W.\ Lemmon,
		\textit{Correlation for the second virial coefficient of water},
		J.\ Phys.\ Chem.\ Ref.\ Data\ \textbf{33}, 369 (2004).

	\bibitem{Walz1997}
		J.Y.\ Walz,
		\textit{Measuring particle interactions with total internal reflection microscopy},
		Curr.\ Opin.\ Colloid Interface Sci.\ \textbf{2}, 600 (1997).

	\bibitem{Dietrich1995}
		S.\ Dietrich and A.\ Haase,
		\textit{Scattering of x-rays and neutrons at interfaces},
		Phys. Rep. \textbf{260}, 1 (1995).

	\bibitem{AlsNielsen2001}
		J.\ Als-Nielsen and D.\ McMorrow,
		\textit{Elements of modern X-ray physics} 
		(Wiley, New York, 2001).

	\bibitem{Jackson1999}
		J.D.\ Jackson,
		\textit{Classical Electrodynamics}
		(Wiley, Hoboken, 1999).    

	\bibitem{Podgornik2016}
		A.\ C.\ Maggs and R.\ Podgornik,
		\textit{General theory of asymmetric steric interactions in electrostatic double layers},
		Soft Matter \textbf{12}, 1219 (2016).	

\end{thebibliography}
\end{document}